\documentclass[11pt,preprint]{aastex}
\usepackage{graphicx}

\begin{document}

\title{THE MILLENNIUM ARECIBO 21-CM ABSORPTION LINE SURVEY. I. \\
TECHNIQUES AND GAUSSIAN FITS}

\author{Carl Heiles}
\affil{Astronomy Department, University of California,
    Berkeley, CA 94720-3411; cheiles@astron.berkeley.edu}

\author{T.H. Troland}
\affil{Department of Physics and Astronomy, University of Kentucky,
Lexington, KY; troland@pa.uky.edu}

\begin{abstract}

	We review the theory of measuring spectral lines in
emission/absorption observations and apply it to a new survey of the
21-cm line against 79 continuum sources.  We develop an observing
technique and least-squares procedure to determine the opacity profile,
the expected emission profile, and their uncertainty profiles.  We
discuss the radiative transfer for the two-component interstellar HI gas
and use Gaussian components, separate ones for the Warm and Cold neutral
media (WNM and CNM), as a practical implementation of a simple but
physically correct model that successfully treats both simple and
complicated profiles.  Our Gaussians provide CNM spin temperatures,
upper limits on kinetic temperatures for both CNM and WNM from the line
widths, column densities, and velocities; we discuss these astrophysical
aspects in Paper II. 

\end{abstract}

\tableofcontents

\section{INTRODUCTION}

	In February 1999 we used the Arecibo\footnote{The Arecibo
Observatory is part of the National Astronomy and Ionosphere Center,
which is operated by Cornell University under a cooperative agreement
with the National Science Foundation.} telescope to begin a series of
Zeeman-splitting measurements of the 21-cm line in absorption against
continuum radio sources. Zeeman-splitting measurements require high
sensitivity and a by-product of this survey is a set of sensitive
emission/absorption line data for 79 sources from which spin
temperatures and other information can be gleaned. 

	In \S \ref{stokespractice} we discuss Arecibo's instrumental
effects and introduce a least-squares technique to account for both them
and for angular derivatives of the HI emission. In \S
\ref{spintempderivation} we discuss the radiative transfer of the
two-component (warm and cold) HI and define our technique of Gaussian
fitting as a practical means to treat radiative transfer in a physically
correct but simple model. \S \ref{gaussiancomps} discusses the practical
implementation of the Gaussian fitting process and the associated
difficulties and uncertainties. \S \ref{slopevsgauss} compares our
method with a previous method for dealing with the radiative transfer,
the ``slope method''. \S \ref{summary} is a brief summary of the paper.
Heiles (2001a) presented a preliminary report of the astrophysical
implications of our Gaussian components on the WNM and CNM; Paper II
(Heiles \& Troland 2002) presents the complete discussion.

\section{EXTRACTING EXPECTED AND OPACITY PROFILES FROM ARECIBO DATA}

\label{stokespractice}

\subsection{ The ON-OFF spectrum}

	In the presence of a continuum source that provides antenna
temperature $T_{src}$, the on-source antenna temperature is

\begin{equation} \label{eqnone}
T_{src}(\nu) = T_{exp}(\nu) + T_{src} e^{-\tau (\nu)} \ ,
\end{equation}

\noindent where $T_{exp}(\nu)$ is the ``expected profile'', which is the
emission that would be observed in the absence of the source, and $\tau
(\nu)$ is the 21-cm line opacity; both of these are functions of
frequency because of the spectral line. The appended symbols $(\nu)$
indicate frequency-dependent quantities within the profile; unappended
temperatures are continuum. In this paper, all temperatures are (Stokes
$I$/2) because they are the average of measurements in orthogonal
polarizations. 

	Consider equation~\ref{eqnone} and assume, for the moment, that
the spatial derivatives are zero. Then the two unknowns
$T_{exp}(\nu)$ and $e^{-\tau(\nu) }$ are easily separated
observationally by taking on-source and off-source measurements, for
which $T_{src}$ changes from zero to the full source intensity. More
generally the spatial derivatives are nonzero; using the on source and
off source measurements, we include spatial derivatives to write a more
complicated version of equation~\ref{eqnone} and subject it to a least
squares analysis, as discussed in detail in \S\ref{lsfitstki}. By this
procedure, the expected emission profile $T_{exp}(\nu)$ and the opacity
profile $e^{-\tau(\nu) }$ (and, in addition, the spatial derivatives)
are well-determined.

\subsection{The Arecibo telescope and electronics}

	The Arecibo telescope has a reflector fixed on the ground and
points by moving the feed structure.  This makes many characteristics of
the beam change as a source is tracked.  These changes are discussed by
Heiles et al (2001a) and documented in more detail on Arecibo's website.
We sampled two linearly polarized channels simultaneously, performing
both auto and crosscorrelations with Arecibo's three-level ``interim''
digital correlator, to generate and instrumentally correct all four
Stokes parameters in the manner discussed by Heiles (2001) and Heiles et
al (2001b). 

	We observed each source by repeating many times an observational
pattern.  A ``pattern'' consists of a sequence of $J$ ``measurements''
at fixed, defined, different sky positions (denoted by subscript $j$)
relative to the source observed for prescribed times.  We designed the
patterns to measure the first and second derivatives of intensity on the
sky so that we could correct for instrumental effects as described
below.  Each pattern contains one or more on-source measurements and at
least four off-source measurements. The total integration time for a
given source consists of $N$ patterns (denoted by subscript $n$).  

	The on-source antenna temperature can vastly exceed the
off-source one.  With three-level correlators it is important to keep
the input level at the optimum value; this meant that the electronics
gains for the on-source measurements (${\cal G}_n$ below) were set lower
than those for the off-source ones ($G_n$ below).  These gains, which
are functions of $n$ but not of $j$, had to be calibrated independently
and have independent uncertainties. 

	For each source we repeated a pattern $N$ times. Each pattern is
characterized by its own gains, so we have $N$ values of ${\cal G}$ and
$G$; we denote individual ones by the subscript $n$, e.g.\ ${\cal G}_n$.

\subsection{The observing technique}

	We have two patterns for observing.  One, called Z4, generates
four off-source emission spectra and the on-source spectrum.  The four
allow determination of the first derivative of the emission line in the
sky.  The other pattern, called Z16, generates sixteen off-source
emission spectra and the on-source spectrum (Figure \ref{crossfig}). 
The sixteen allow determination of not only the first derivative of the
emission line in the sky, but all second derivatives.  The following
details must be accounted for in the data reduction: \begin{enumerate}

\item The on-source electronics gain ${\cal G}_n$ for each pattern
$n$ is not perfectly calibrated. 

\item The off-source electronics gain $G_n$ for each pattern $n$ is not
perfectly calibrated. 

\item The on-axis gain (``Kelvins per Jansky'') $[{\rm K \ Jy}^{-1}]_n$
of the telescope changes with $(az,za)$, so that the source deflection
changes. This quantity is known from previous calibration data; we adopt
it from a table lookup.

\item The off-source cold-sky system temperature $T_{R,n}^*$ changes
with $za$ (but not by much). Our notation, with the subscript $R$,
implies ``Receiver'' temperature, but this is not quite correct. Rather,
it is the off-source cold-sky system temperature, which is the actual
receiver temperature plus other sources of cold-sky noise including,
specifically, the cosmic background radiation and also the
$za$-dependent contribution from the ground; $T_{R,n}^*$  depends on
$(az,za)$ and is known from previous calibration data. We do not measure
it explicitly, but rather adopt it from a table lookup. 

\item Arecibo's beam shape changes with $(az,za)$.  In particular, the
location of the first zero in the antenna beam response changes and,
moreover, does not always exist.  This means that the off-source spectra
contain some remnant of the source intensity so one cannot obtain the
expected profile by going just a little way off source. 

\end{enumerate}

\subsection{ Obtaining the continuum source flux and antenna temperatures}
\label{continuumcal}

	In this and future sections we need to distinguish between
measured qauantities and true quantities. The difference arises from the
fact that the gain calibrations ($G$ and $\cal G$) are imperfect.  The
measured quantities are denoted by the superscript $m$ and the true
quantities by $*$. For system temperatures (equations \ref{eqntwoz}
below), the measured and true quantities are related by gain factors in
the sense that the gain factor multiplied by the true quantity equals
the measured one. Antenna temperatures are derived from system
temperatures by subtracting a table-lookup receiver temperature
(equations \ref{srctempz} below), so the relationship is more
complicated. Measured quantities also have random noise, but in the
following we temporarily assume they are noise-free so that we can write
(schematically) $T_{sys}^m = G T_{sys}^*$. 

	For a particular off-source position $j$ within pattern $n$, we
directly measure the system temperature $T_{sys,n,j}^m$, which is

\begin{mathletters} \label{eqntwoz}
\begin{equation} \label{eqntwoa}
T_{sys,n,j}^m = G_n T_{sys,n,j}^*
\end{equation}

\noindent and, correspondingly for the on-source position (denoted by $j=s$), 

\begin{equation}
T_{sys,n,s}^m = {\cal G}_n T_{sys,n,s}^* 
\end{equation}
\end{mathletters}

\noindent The off-source antenna temperature, i.e.\ the excess response
of the telescope over cold sky, includes the response to the source
(which might enter through sidelobes) as well as any diffuse emission
that happens to lie in the same direction and is simply

\begin{equation}
T_{ant,n,j}^* = T_{sys,n,j}^* - T_{R,n}^*  \ . 
\end{equation}

\noindent We define the measured antenna temperatures as

\begin{mathletters} \label{srctempz}
\begin{equation} \label{srctempa}
T_{ant,n,j}^m \equiv T_{sys,n,j}^m - T_{R,n}^* = 
	G_n T_{sys,n,j}^* - T_{R,n}^* \  
\end{equation}

\noindent with the corresponding equation for the on-source position,
for which  we define the on-source antenna temperature to be the source
temperature $T_{src,n}$: 

\begin{equation} \label{srctemp}
T_{src,n}^m \equiv T_{sys,n,s}^m - T_{R,n}^* = 
	{\cal G}_n T_{sys,n,s}^* - T_{R,n}^* \ .
\end{equation}
\end{mathletters}

\noindent We derive the source flux by assuming that all of the antenna
temperature, including that portion that comes from unrelated diffuse
emission, arises from the source, i.e.\ we define

\begin{equation} \label{fluxdefinition}
S_{src} \equiv {T_{src,n}^* \over [{\rm K \ Jy}^{-1}]_n}
	= { {T_{sys,n,s}^m \over {\cal G}_n}   - T_{R,n}^* 
	\over [{\rm K \ Jy}^{-1}]_n}
\end{equation}

\noindent Except for ${\cal G}_n$, the quantities on the right hand side
are known from our data or from previous calibration.  The gains ${\cal
G}_n$ are randomly distributed with $|{\cal G}_n - 1| \ll 1$, so $1
\over {\cal G}_n$ is also randomly distributed.  Therefore, from the set
of $N$ patterns we obtain the best estimate for $S_{src}$ from a
least squares fit of the $N$ equations

\begin{equation}
[{\rm K \ Jy}^{-1}]_n S_{src} = T_{sys,n,s}^m - T_{R,n}^*  \ ,
\end{equation}

\noindent i.e., by assuming all ${\cal G}_n=1$. 

	Having performed the least squares fit, we rewrite equation
\ref{fluxdefinition} to solve for the unknown ${\cal G}_n$:

\begin{equation}
{\cal G}_n = {T_{sys,n,s}^m \over
[{\rm K \ Jy}^{-1}]_n S_{src} + T_{R,n}^* } \ .
\end{equation}

\noindent All quantities on the right-hand side are known, so this
explicitly provides the value for each ${\cal G}_n$ for each pattern,
allowing us to correct the individual on-source measurements
$T_{sys,n,s}^m$ for the gain error (but not the off-source measurements;
see \S \ref{lsfitstki}). 

\subsection{ The meaning of our derived source fluxes}

	We derive the source flux $S_{src}$ with equation
\ref{fluxdefinition}. In essence, this averages the difference between
the ${\cal G}_n$-corrected system temperatures $T_{sys,n,s}^m$ and the
predicted cold-sky antenna temperatures $T_{R,n}^*$.  Our sources are
all small, no more than a few arcmin in diameter.  If a source lies
within an extended region of emission, then our derived source flux
includes the antenna temperature from that extended region.  Thus, our
source fluxes should be systematically larger than those measured
interferometrically.  This does not affect the scales of our derived
opacity spectra because they are derived from ON--OFF differences. 

\subsection{ Least-squares fitting the spectra}
\label{lsfitstki}

\subsubsection{The least-squares technique: theory} \label{lsfitstkithy}

	Conventionally, one observes off-source positions to obtain the
``expected profile'' $T_{exp}(\nu )$, which is the line profile one
would observe at the source's position if the continuum source were
turned off. One uses the expected profile for two purposes: one
subtracts it from the on-source spectrum, which difference provides the
opacity spectrum; and one combines it with the opacity to obtain the
spin temperatures. 

	This conventional technique is not perfect because the
off-source emission spectra differ from the expected profile. There are
two reasons. First, the antenna response to the continuum source is not
zero for the off positions, so the off-source emission spectra are
contaminated by a small, unknown contribution from the opacity spectrum.
Second, there is angular structure in emission spectra. We treat these
problems using a least squares technique.

\begin{figure}[h!]
\begin{center}
\includegraphics[width=3.5in] {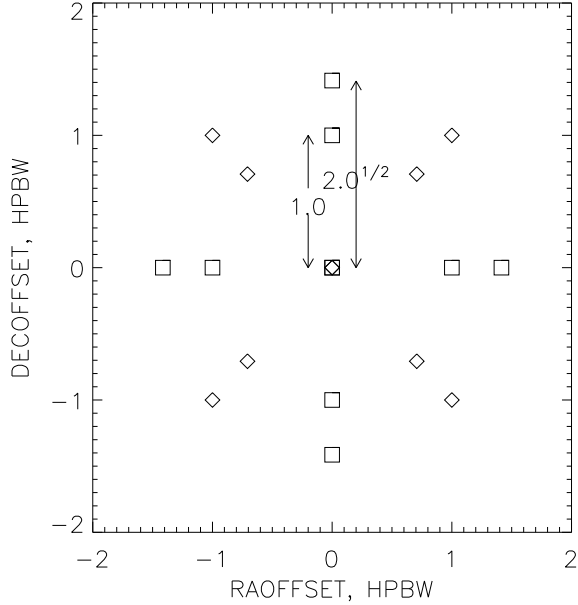} 
\end{center}

\caption{The 17-point measurement grid of the Z16 pattern. It consists
of two crosses, one aligned with ra/dec (squares) and one at $45^\circ$
to ra/dec (diamonds).  The innermost points are $1.0 \, H\!P\!B\! W$ and
the outermost points $2^{1/2} H\!P\!B\!W$ from the central point. 
\label{crossfig}} \end{figure}

	We assume that the angular structure in each spectral channel
can be represented by a Taylor series expansion; for our early data we
carried it to first order and we obtained the derivatives from a
four-point off-source grid centered on the source (the Z4 pattern). 
Later we carried the expansion to second order.  Obtaining second
derivatives from measurements on the two-dimensional sky requires
measuring a minimum of nine independent positions.  Normally one selects
positions on a nine-point grid centered on the position of interest, and
includes the central position as an equal partner in the calculation of
the derivatives.  However, in this case the central point is not an
equal partner because it contains the continuum source.  Therefore we
need at least one additional point.  Moreover, we want redundancy so
that we can estimate the uncertainties in the derived quantities.  In
practice, we made measurements on the 17-point grid shown in
Figure~\ref{crossfig}; there is one on-source point and 16 off-source
points (the Z16 pattern).

	As described above, for each pattern we corrected the on-source
measurement for the gain errors ${\cal G}_n$.  However, all the
off-source measurements for pattern number $n$ have a single
independently-calibrated gain $G_n$ which differs from ${\cal G}_n$.  If
there were no gain errors for the off-source measurements, i.e.~if $G_n
= 1$, then each measurement $j$ within pattern $n$ would provide the
true system temperature $T_{sys,n,j}^*(\nu)$ at each frequency channel
$\nu$ in the profile, which satisfies the equation

\begin{equation} \label{ls1}
T_{sys,n,j}^*(\nu) - T_{R,n}^* = [T_{exp}(\nu)] +
\left[ \partial T_{exp}(\nu) \over \partial \alpha \right] \Delta\alpha_{j} + 
\left[ \partial T_{exp}(\nu) \over \partial \delta \right] \Delta\delta_{j} + $$
$$
\left[ \partial T_{exp}(\nu)^2 \over \partial^2 \alpha \right]
     {(\Delta\alpha_{j})^2 \over 2} + 
\left[ \partial T_{exp}(\nu)^2 \over \partial\alpha \partial\delta \right]
     (\Delta\alpha_{j}) \ (\Delta\delta_{j}) + 
\left[ \partial T_{exp}(\nu)^2 \over \partial^2 \delta \right]
     {(\Delta\delta_{j})^2 \over 2} + 
[e^{-\tau (\nu)}] T_{ant,n,j}^* \ .
\end{equation}

\noindent Here, as before, we have $N$ patterns, each of which is
denoted with subscript $n$, and within each pattern we have $J$
measurements, each of which is denoted by subscript $j$. Each pattern
consists of measurements at $J=17$ positions, one of which is directly
on-source; even the off-source positions have nonzero $T_{ant,n,j}^*$,
as discussed above.  The appended symbols $(\nu)$ indicate
frequency-dependent quantities within the profile; unappended
temperatures are continuum. Quantities to be derived from a least
squares treatment are enclosed in square brackets.  There are 7 such
quantities for the full second-order expansion and 4 for the first-order
one. 

	In fact, however, we have off-source gain uncertainties, just as
we had on-source gain uncertainties in \S\ref{continuumcal}.  To convert
from the idealized starred quantities to the measured ones, we multiply
both sides of equation \ref{ls1} by the off-source gain factor $G_n$ and
use equations \ref{eqntwoa} and \ref{srctempa}, yielding the more complicated 

\begin{equation} \label{ls2}
T_{sys,n,j}^m(\nu) - T_{R,n}^* = [G_n] \left\{ [T_{exp}(\nu)] +
\left[ \partial T_{exp}(\nu) \over \partial \alpha \right] \Delta\alpha_{j} + 
\left[ \partial T_{exp}(\nu) \over \partial \delta \right] \Delta\delta_{j} + 
\right. $$
$$
\left. \left[ \partial T_{exp}(\nu)^2 \over \partial^2 \alpha \right]
     {(\Delta\alpha_{j})^2 \over 2} + 
\left[ \partial T_{exp}(\nu)^2 \over \partial\alpha \partial\delta \right]
     (\Delta\alpha_{j}) \ (\Delta\delta_{j}) + 
\left[ \partial T_{exp}(\nu)^2 \over \partial^2 \delta \right]
     {(\Delta\delta_{j})^2 \over 2} \right\} + $$
$$ + [e^{-\tau (\nu)}] (T_{sys,n,j}^m - 
   [G_n]T_{R,n}^*) + ([G_n] -1)T_{R,n}^* \ .
\end{equation}

	The additional complexity of this equation does not lie simply
in its being more cumbersome and also requiring a nonlinear, instead of
a linear, least-squares fit.  Rather, it means that we have $N$
additional quantities, the $G_n$, for a total of ($7+N)$ unknowns for
each spectral channel.  We cannot solve this on a channel-by-channel
basis because $G_n$ is identical for all channels and, moreover, for a
single channel the equations are degenerate. To remove the degeneracy we
need to have different system temperatures, and these are provided by
different channels with their varying 21-cm line emission and absorption
spectral values. 

	We can solve the system of equations by considering all
channels simultaneously in the fit to the $N$ patterns.  If $C$ is
the number of channels in the profile, this means solving $CNJ$
equations.  For each channel there are 7 unknowns [$\tau(\nu)$ together
with $T_{exp}(\nu)$ and its various derivatives] and for the ensemble of
patterns there are $N$ unknown gains, so the total of unknowns is
$(7C + N)$.  The number of equations exceeds the number of
unknowns and the solution exists as long as there is no degeneracy.

	However, with current computers it is difficult (but hardly
impossible) to solve these $CNJ$ equations using least squares because
often $C \sim 1000$ and typically $N \sim 100$; the entire assembly
constitutes $CNJ \sim 1.5 \times 10^6$ independent measurements with
$\sim (7000 + N)$ unknowns, so using matrix techniques requires
generating and manipulating a $\sim (1.5 \times 10^6) \times 7100$
matrix, which occupies $\sim 40$ Gb of memory. 

	We attack the problem by splitting it into two parts; this is
motivated by the fact that typically all $G_n = 1$ to within a few
percent. First, we assume all $G_n = 1$ and solve equation \ref{ls1} for
the spectral quantities, treating each channel individually, first
setting $T_{ant,n}^* = T_{sys,n}^m - T_{R,n}^*$. Solving these $NJ \sim
1500$ separate equations by least-squares provides the 7 unknowns for
each channel; repeating the process for each channel provides all of the
$\sim 7000$ unknowns. Next we use these derived quantities in a
least-squares fit for the $N$ $G_n$, using all frequency channels
simultaneously: this fit has, as before, $\sim 1.5 \times 10^6$
measurements, but only $N \sim 100$ unknowns, and is tractable. This
provides the $N$ gains; using these, we again perform the solution for
the spectral quantities for each channel individually. It is unnecessary
to iterate further, because this procedure converges rapidly and a
single iteration derives the gains to high accuracy.

	There is one additional complication in the least-square
fitting, which is that each of the $N$ patterns, each of the $J$
measurements within each pattern, and each frequency channel have a
different intrinsic noise.  For each of the $NJ$ measurements the system
temperature is different; in particular, the on-source measurements
within a pattern have a significantly, sometimes much, higher system
temperature than the off-source measurements.  Moreover, the system
temperatures depend on frequency; for example, at the peak optical depth
of the 21-cm line, the on-source system temperature is smaller than that
off the line. Finally, some of our data were taken with the five-point
observing grid and some with the 17-point grid; the individual
measurements for these grids have different integration times. The
equation of condition for each measurement for a particular spectral
channel received the weight $W_n(\nu)$ equal to the reciprocal of the
rms noise $\sigma(T)$ calculated in the usual way\footnote{ see \S15.4
of Press et al (1992), which calls the set of equations of condition the
``design matrix''.}

\begin{mathletters}
\begin{equation}
W_n(\nu) = {1 \over \sigma(T(\nu))} 
\end{equation}

\noindent where
\begin{equation}
\sigma(T(\nu) ) = {0.77 T_{sys}(\nu) \over \sqrt{t \Delta \nu} } 
\end{equation}
\end{mathletters}

\noindent where $T_{sys}(\nu)$ is the system temperature in that
channel, $t$ the integration time, and $\Delta \nu$ the frequency
separation between channels; we empirically determined the factor 0.77
as being what was required to reproduce the observed $\sigma(T(\nu))$. 

\subsubsection{The least-squares technique: illustration}
\label{lsfitstkiobs}

\begin{figure}[h!]
\begin{center}
\leavevmode
\plottwo{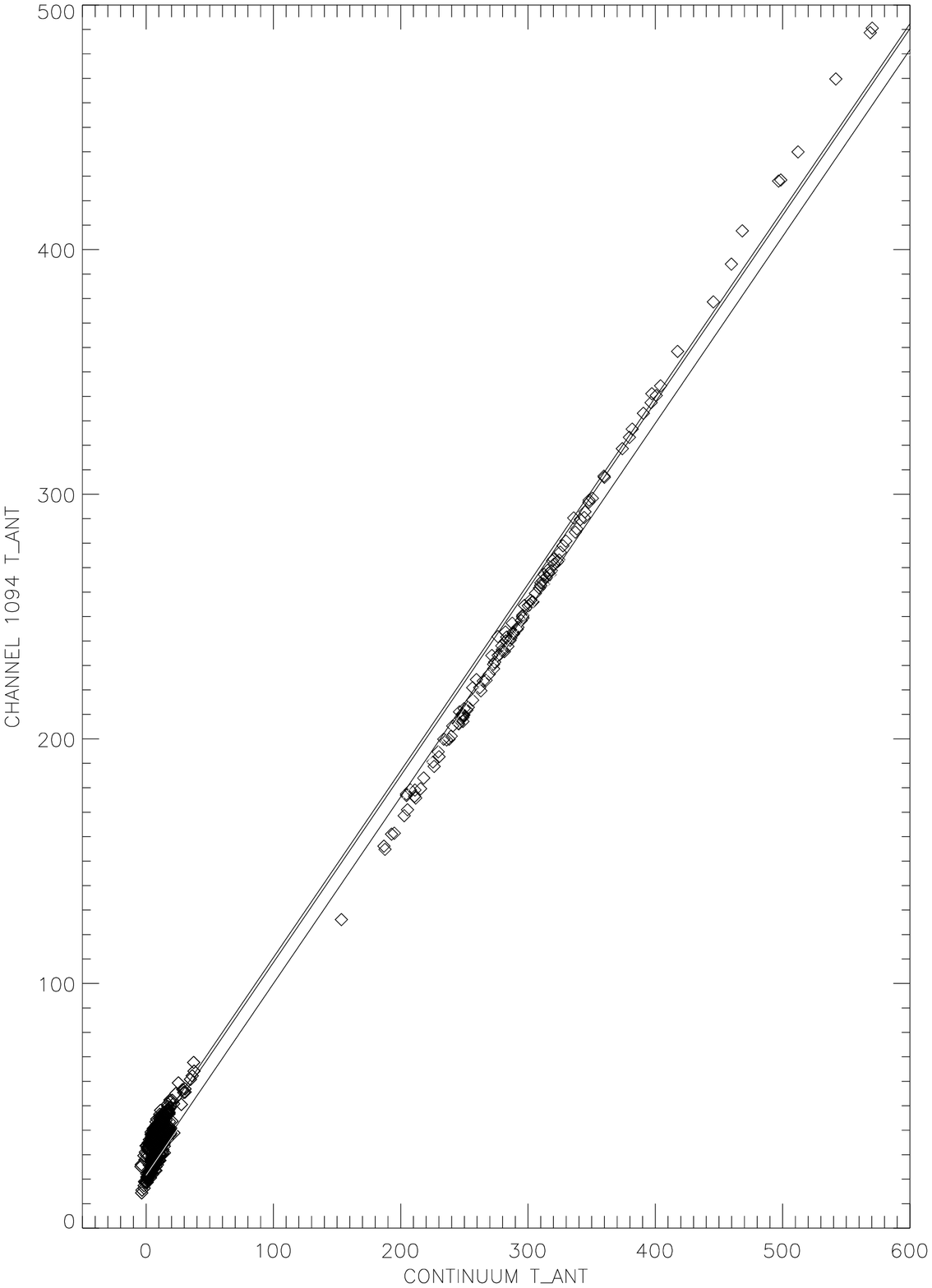}{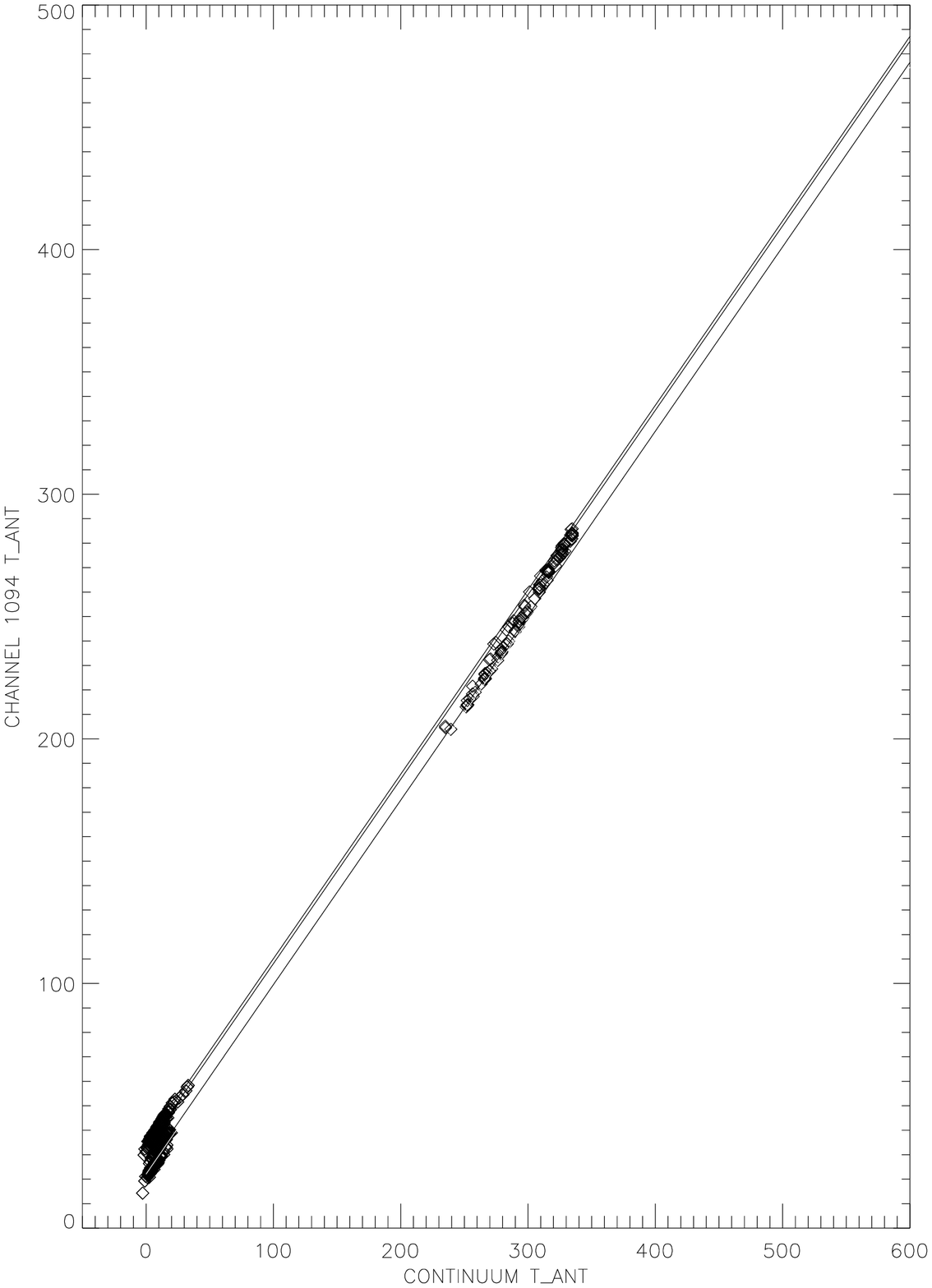}
\end{center}

\caption{$T_{ant}$ for frequency $\nu 1094$ (spectral channel 1094)
versus $T_{ant}$ for the continuum.  On-source datapoints for
pattern $n$ are those having $T_{ant,n} > 100$ K; the others are
off-source points, four for each pattern.  The left panel shows the
measured values without gain correction, the right panel with gain
correction.  See \S \ref{lsfitstkiobs}.  \label{gainsfig}} \end{figure}

	Figure \ref{gainsfig} illustrates the fit technique and results
for 3C454.3.  We illustrate using this source for two reasons.  First,
it is strong, so that the off-source positions have an easily detectable
continuum flux from the source.  In addition, it has unusually large
gain errors, both for on-source ${\cal G}_n$ and off-source $G_n$, for
the following reason.  It is a VLBI calibrator and was observed near
Solar maximum in early March 1999, when it had the same right ascension
of the Sun; this put it only about $23^\circ$ from the Sun.  It suffered
severe interplanetary scintillation, which produced some variability in
its apparent intensity on the time scale of a few seconds, which was the
interval during which our cal was turned on and off; these variations
produced significant errors in the measured cal deflection, which led to
gain errors for the on-source spectra as high as $\sim 50\%$ and for the
off-source spectra as high as $\sim 10\%$. 

	Figure \ref{gainsfig} plots the measured antenna temperatures at
frequency $\nu 1094$ (spectral channel 1094) versus the measured
continuum antenna temperatures.  We chose this channel because it has the
peak optical depth. The vertical scale of the left panel is $T^m_{ant,
n, j}(\nu1094) = T^m_{sys, n, j}(\nu1094)- T^*_{R,n}$, i.e.\ the
spectral antenna temperature {\it not} corrected for gain. The vertical
scale of the right panel is the gain-corrected equivalent $T^*_{ant, n,
j}(\nu1094)$, where we use ${\cal G}_n$ and $G_n$ to correct the
on-source and off-source measurements, respectively. 

	These plots are for the Z4 pattern with one on-source and four
off-source measurements.  The on-source datapoints are those having
$T_{ant,n} > 100$ K.  There is one on-source datapoint for each pattern
$n$; its gain is the on-source gain ${\cal G}_n$.  There are four
off-source datapoints for each pattern $n$; their gains are all
identical, equal to the off-source gain $G_n$. 

	There are four plotted lines in each of the two panels of Figure
\ref{gainsfig}. Each line is a least-squares fit for the measurements of
the on position and one of the four off positions.  They have slightly
different $y$-intercepts but, as a constraint of the fits, identical
slopes.  The slope is $e^{-\tau(\nu1094)}$.  The four $y$-intercepts are
the observed antenna temperatures for the four off positions, averaged
over all $N$ patterns. 

	In the left panel the dispersion of on-source antenna
temperatures is huge, which is a result of the anomalously high gain
errors produced by the scintillation.  The dispersion of the off-source
points is much less, but still larger than expected.  These excessive
dispersions are produced by scintillation.  The dispersions are much
smaller for the gain-corrected data in the right panel. 

	In the right panel, the corrected on-source antenna temperatures
disperse along the plotted line.  This shows that the gain-corrected
dispersion does not arise from statistical noise.  Rather, it is real
and reflects Arecibo's $(az,za)$-dependent point-source gain $[{\rm K \
Jy}^{-1}]_n$.  Similarly, the off-source points disperse along the line;
this is produced by the small responses of the beam to the source at the
off-source positions, which differ from one pattern $n$ to another as
the telescope beam rotates on the sky. 

\subsection{ Evaluation of instrumental effects }
\label{realerrors}

	Our ultimate goal is to derive only two of the
spectral profile parameters in equations \ref{ls1} and \ref{ls2}, namely
the expected profile $T_{exp}(\nu )$ and the opacity profile $e^{-\tau
(\nu )}$. We need the other profile parameters involving angular
derivatives only for the purpose of estimating uncertainties in the two
desired profiles.

	The least squares process itself directly provides 
uncertainties in the $M$ derived parameters; the equation for the $m$'th
parameter is, in schematic form,

\begin{equation}
error_m = C_{mm} {\sum_{n=0}^{N-1} residual_n^2 \over N (N-M)}
\end{equation}

\noindent where $C_{mm}$ is the $m$'th diagonal element of the
covariance matrix and $N$ is the number of independent patterns. 
There are three basic assumptions in this equation.  Most important is
that the measurements are independent.  Also, that the residuals are
randomly (normally) distributed about the true value and that there is
no covariance among the derived quantities. 

	However, this equation cannot apply for our derived spectral
parameters: specifically, while we can divide by $(N-M)$ we should
certainly not divide by the additional factor of $N$.  The reason is
that we have a number $N$ identical patterns, meaning $N$ measurements
of each OFF position and the ON position.  Suppose that there were no
instrumental noise; then the residuals for all patterns would be
identical, because the residuals would produced be only by the higher
angular derivatives than we measure.  In this case, dividing by the
additional factor of $N$ produces a meaningless and unrealistic
reduction in $error_m$.  In fact, each of the $N$ patterns does have
some instrumental noise.  We need a technique to average the
instrumental noise but not to artifically decrease the derived
$error_m$. 

	We describe our technique for deriving the error in
$T_{exp}(\nu)$ for the 17-point pattern; we use a similar technique for
the 5-point pattern.  The 17-point pattern has 16 OFF measurements.  For
a set of $N$ patterns, we average each OFF measurement over the $N$
patterns; this averages the instrumental noise and provides a fairly
noise-free measurement of the antenna temperature spectrum at each of
the 16 independent OFF positions.  Also, equation \ref{ls1} provides a
{\it prediction} of the same spectrum at each OFF position once the
fitting process has yielded spectra for the unknowns in that equation. 
For each channel independently, we calculate the 16 differences between
the measured and predicted antenna temperatures.  These 16 residuals
represent 16 different estimates of the error of the fit in predicting
the antenna temperature in that channel.  We average the squares of
these 16 residuals and take the square root, obtaining an rms spectrum
$\Delta T (\nu )$ in which the spectral values are statistically
representative {\it uncertainties} of the predicted antenna temperatures
at the OFF positions.  $\Delta T (\nu )$ should also characterize the
uncertainties of the predicted antenna temperatures at the ON position;
thus $\Delta T_{exp} (\nu ) = \Delta T (\nu )$. 

\begin{figure}[p!]
\begin{center}
\includegraphics[height=6in] {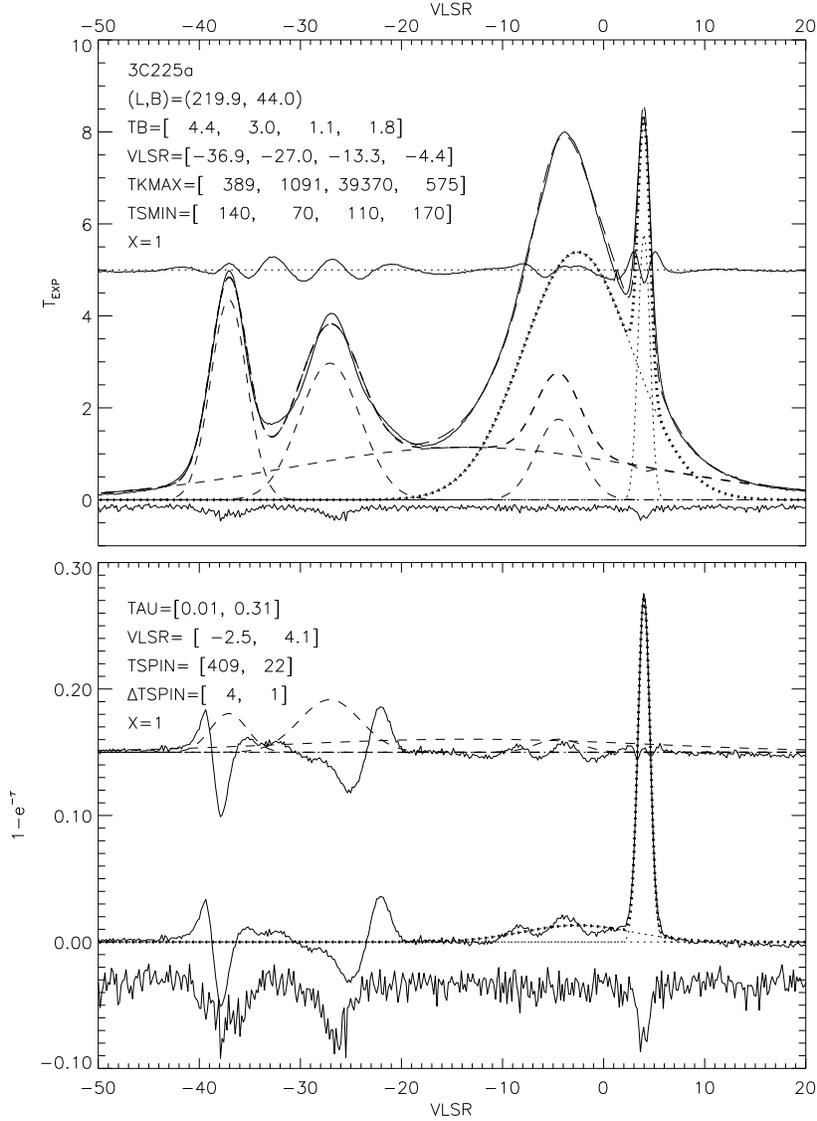} 
\end{center}

\caption{Illustrating the derived spectral parameters and their
uncertainties as derived in \S\ref{realerrors} for the source 3C225a,
which has severe problems with derivatives.  In the top panel, the solid
line profile extending above zero is the expected profile $T_{exp}(\nu)$
and the solid line profile extending below zero is the (negative of) its
uncertainty $\Delta T_{exp}(\nu)$.  In the bottom panel, the solid line
profile extending above zero is the opacity profile $[1 -
e^{\tau(\nu)}]$ and the solid line profile extending below zero is the
(negative of) its uncertainty $\Delta e^{-\tau(\nu)}$.  Gaussian
parameters for the WNM (top) and CNM (bottom) are given in square
brackets.  For a complete description of other features in these
two-panel plots, see \S \ref{twopanelplots} and \ref{tabular}.  Plots
for all sources are in the electronic edition of {\it The Astrophysical
Journal}. Also, they can be retreived by ftp from
vermi.berkeley.edu/pub/zobs/bigfig.ps .  \label{zec20plot}} \end{figure}

	The error in the expected profile is simply $\Delta
T_{exp}(\nu)$.  To derive the error in the optical depth profile, we
recall that if the source were turned off (making its flux $S = 0$ Jy)
then the ON-source antenna temperature $T_{ON, \, 0 Jy}$ is the expected
profile

\begin{mathletters}
\begin{equation}
T_{ON, \, 0 Jy} (\nu ) = T_{exp} (\nu )
\end{equation}

\noindent and with the source turned on ($S = S_{src}$ Jy) we have

\begin{equation}
T_{ON, \, S_{src}Jy} (\nu ) = T_{exp} (\nu ) + T_{src} e^{-\tau (\nu )}
\end{equation}

\noindent where $T_{src} = [{\rm K \ Jy}^{-1}] S_{src}$ is the continuum
antenna temperature of the source.  Combining these provides

\begin{equation}
e^{-\tau (\nu )} = {{T_{ON, \, S_{src}Jy} (\nu ) - T_{exp} (\nu )} 
  \over T_{src}} 
\end{equation}

\noindent To calculate the uncertainty in the opacity profile $\Delta
e^{-\tau (\nu )}$, we retain only the uncertainty in $T_{exp}( \nu )$; the
fractional errors in $T_{ON, \, S_{src}Jy} (\nu )$ and $T_{src}$ are
negligible because they are averages of $N$ directly-measured
quantities. This gives 

\begin{equation} \label{deltatau}
\Delta e^{-\tau (\nu )} = {\Delta T_{exp} (\nu ) \over T_{src} }
\end{equation}
\end{mathletters}

	Figure \ref{zec20plot} illustrates these results for 3C225a, for
which the angular derivatives for $V_{LSR} \lesssim -15$ km s$^{-1}$ 
are particularly deleterious. The top panel shows the expected profile
$T_{exp}(\nu)$ and its error $\Delta T_{exp}(\nu)$; the bottom panel,
the opacity profile $e^{-\tau (\nu )}$ and its error $\Delta e^{-\tau
(\nu )}$.  The relative error in the expected profile $\Delta
T_{exp}(\nu)/ T_{exp}(\nu)$ is small, which is typical for all sources. 

	For our sources the relative errors in the opacity profile are
usually, but not always, small.  For the example of 3C225a the errors
for the negative velocity peaks are comparable to the measured values. 
This is fortunate, as the measured values for $e^{-\tau (\nu)}$ exceed
unity---meaning $\tau < 0$, which implies an interstellar maser! We
don't expect negative spin temperatures for the 21-cm line, and our
error analysis shows that these measurements are meaningless.  This
source, 3C225a, is a serious example of large angular derivatives in the
21-cm line emission.  Some sources are equally serious, although none of
these others happens to exhibit apparent maser emission as a result. 

\subsection{Machine-readable data and the two-panel plots} \label{twopanelplots}

	Table \ref{fulldatatable} presents the expected profile, the
opacity profile, and the various derivatives of the emission profile for
all sources. 

	We exhibit our results for each source in a plot with two
panels.  We present plots for all sources in the electronic edition of
{\it The Astrophysical Journal}, of which Figure \ref{zec20plot} is an
example.  For each source, we choose the smallest possible velocity
range for each plot because we are interested in the fine-scale velocity
structure.  Sometimes this requires us to cut off some emission and/or
opacity components that lie outside the displayed range; however, we do
this only when the fits outside the displayed range are good.  In no
case do we restrict the velocity range when the fits do not agree with
the data, or if there is interesting profile structure, outside the
range.  For two sources that have both wide- and narrow-velocity
structure we present two velocity ranges. 

	In the top panel, the solid line profile extending above zero is
the expected profile $T_{exp}(\nu)$.  The dashed curves show the fitted
emission from each individual WNM component, including absorption by the
CNM (described in \S\ref{spintempderivation}), and the heavy-dashed
curve shows the totality of WNM fitted Gaussians $T_{B,WNM}$ (equation
\ref{gauss3}).  The solid line near mid-profile-height is the residuals
of the data from the fit.  The dotted curves show the intrinsic emission
from each individual CNM component, not including absorption from the
CNM components that lie in front, and the heavy dots the predicted
emission of all CNM components including absorption from the CNM
components that lie in front ($T_{B,CNM}$ in equation \ref{gauss2}). 
The long-dashed curve is the predicted expected profile $(T_{B,WNM} +
T_{B,CNM})$.  The solid line profile extending below zero is the
(negative of the) uncertainties $\Delta T_{exp}(\nu)$ described in
\S\ref{realerrors}. 

	In the bottom panel, the solid line profile extending above zero
is the opacity profile $[1 - e^{\tau(\nu)}]$.  The dotted curves show
each individual CNM opacity Gaussian and the heavy dots show the sum of
all the CNM Gaussians $[1 - e^{- \tau (\nu)}]$, with $\tau (\nu)$ from
equation \ref{gauss1}.  The solid line near mid-profile-height is the
residuals of the data from the fit.  The dashed curve near mid-profile
height shows the sum of the lower limits of WNM opacity (\S
\ref{wnmproperties}), separately for each WNM component. The solid line
profile extending below zero is the (negative of the) uncertainties in
the opacity profile $\Delta e^{-\tau(\nu)}$ (equation \ref{simply}). 

	Annotations in the top and bottom panels list properties of the
WNM and CNM Gaussian components, respectively, listed in order of
increasing velocity.  In the upper panel, TB is the fitted intrinsic,
unabsorbed brightness temperature of each WNM component ($T_{0,k}$ in
equation \ref{gauss3}) and TKMAX is the upper limit kinetic temperature
from the linewidth (\S \ref{tabular}).  TSMIN is the lower limit on WNM spin
temperature (\S \ref{wnmproperties}).

	In the lower panel, TAU is the peak optical depth of each CNM
component $\tau_{0,n}$ in equation \ref{gauss1}, TSPIN is the fitted
value for spin temperature $T_{s,n}$ in equation \ref{gauss2}, and
$\Delta$TSPIN is the uncertainty in $T_{s,n}$ (equation \ref{simply}). 

	Also, each panel has a parameter $X$, which is the factor by
which both the residuals and profile errors are magnified for clarity. 
If the bottom panel has no properties listed, then no CNM components
were fit. 

\subsection{Comparison of our results with previous ones} \label{dstpst}

	We compare our results with two of the most sensitive previous
HI line absorption surveys, Dickey, Salpeter, \& Terzian (1978; DST) and
Payne, Salpeter, \& Terzian (1982; PST). The comparisons cannot be exact
because we don't have the original data. We have 10 sources in common
with PST; all show reasonable agreement.  For a few we see weak opacity
tails where PST might not and for a few the scales of the measured
quantity, which is $(1 - e^{-\tau (\nu)})$, differ by up to $\sim 15\%$.

	We have 15 sources in common with DST, of which 13 have
detectable absorption lines. Of these 13, most show reasonable
agreement, again with possible differences in profile wings and scale. 
Three agree badly: 3C237, 3C310, and 3C348. The disagreements are in the
line shape: the DST profiles are either broader or have more components
than ours. Profiles with more components appear like ``ghosts'', with a
pattern of less-intense components shifted with respect to the same
pattern of more intense ones. PST observed 3C237, which result agrees
with ours. Dickey (personal communication) confirms our suspicion that
the local oscillator during some of his observations was either unstable
or improperly set for Doppler correction because of the transient nature
of the Arecibo software at the time.  Greisen and Liszt (1986) observed
3C348 interferometrically and obtained an opacity spectrum that agrees
with ours; they erroneously ascribed their disagreement with DST to
difficulties with the single-dish observing and data-reduction process. 

\section{ DERIVING CNM SPIN TEMPERATURES AND WNM BRIGHTNESS
TEMPERATURES: RADIATION TRANSFER IN THE 21-cm LINE}
\label{spintempderivation}

\subsection{ A single temperature HI}

	Much HI emission/absorption work has assumed a single
temperature HI component, for which the equation of transfer is

\begin{equation} \label{singletempeqn}
T_{exp}(\nu)= T_s (1 - e^{-\tau (\nu)})
\end{equation}

\noindent Both the expected profile $T_{exp}(\nu)$ and the opacity
profile $\tau (\nu)$ are measured quantities, so one obtains the spin
temperature directly from this equation.

	However, the assumption of a single temperature is clearly
inadequate because emission profiles are always wider than opacity
profiles. This means that some HI is so warm that it produces emission,
contributing to $T_{exp}(\nu)$  but having negligible opacity; this
produces spin temperatures $T_s$ that lie between the warm and cold
values.

\subsection{ A two-component HI} \label{mebold}

	This observational situation leads to the concept of a
two-component medium, with the Cold Neutral Medium (CNM) providing the
opacity and also brightness temperature and the Warm Neutral Medium
(WNM) providing only brightness temperature. Separating the CNM and WNM
contributions to $T_{exp}(\nu)$ then becomes the major difficulty. 

	The equation of transfer for this case is discussed recently by
Mebold et al (1997) and Dickey et al (2000). The expected profile
consists of the WNM and the CNM components:

\begin{equation}
T_{exp}(\nu)= T_{B,WNM}(\nu) + T_{B,CNM}(\nu)
\end{equation}

\noindent Here a fraction $\cal F$ of the WNM lies in front of of the
CNM so that a fraction $(1 - {\cal F})$ of the WNM emission is
absorbed by the CNM gas: 

\begin{equation} \label{intrinsictb}
T_{B,WNM}(\nu) = T_{E,WNM}(\nu) [ {\cal F} + (1 - {\cal F})e^{-\tau (\nu)}]
\end{equation}

\noindent where $T_{E,WNM}(\nu)$ is the intrinsic, unabsorbed emission
from the WNM gas and $\tau(\nu)$ is the CNM opacity. The emission from
the CNM is just the usual

\begin{equation}
T_{B,CNM}(\nu) = T_s (1 - e^{-\tau (\nu)})
\end{equation}

These authors discuss and illustrate how to separate the components by
plotting the two observables $T_{exp}(\nu)$ and $[1 - e^{-\tau (\nu)}]$
against each other:

\begin{equation} \label{slopeeqn}
T_{exp}(\nu) = T_{E,WNM}(\nu) + 
	[ T_s - T_{E,WNM}(\nu) (1 - {\cal F})] [1-e^{-\tau (\nu)}] 
\end{equation}

\noindent The $y$-intercept is $T_{E,WNM}(\nu)$ and the slope is  $[ T_s
- T_{E,WNM} (1 - {\cal F}])$. Because $\cal F$ lies between 0 and 1, the
slope provides limits on $T_s$. Some inaccuracy occurs because in the
region where the slope is defined the quantities change with $\nu$, but
nevertheless the method seems to produce reasonable results. We call
this the ``slope method''. 

        Previous single-dish authors, in contrast, implicitly assume
that clouds are {\it not} isothermal.  They derive the spin temperature
from equation \ref{singletempeqn} at the opacity profile peak, where the
derived temperature is coldest;  this spin temperature lies between
between the lowest and highest temperatures along the line of sight
through the cloud. Moreover, temperatures derived off of the opacity
profile peak are warmer.  Thus each point on their histograms does not
represent a true mean cloud temperature; rather, it represents the
lowest derived temperature, which itself is a weighted mean temperature
for which the weights are unknown.

\subsection{ Gaussian components and radiative transfer} \label{gaussrad}

	Our emission and opacity profiles clearly consist of multiple
components, which often overlap and have different spin temperatures.
This requires generalizing the equations in \S\ref{mebold}, which we
accomplish by assuming the components can be represented as Gaussians. 

        For each CNM opacity spectrum, we represent its components'
optical depths by a set of $N$ Gaussians. Thus we least-squares fit the
observed opacity spectrum $e^{- \tau(\nu)}$, where

\begin{equation} \label{gauss1}
\tau(\nu) = \Sigma_0^{N-1} \tau_{0,n} e^{-[(\nu-\nu_{0,n})/ \delta
\nu_n]^2} \ 
\end{equation} 

\noindent in which $(\nu_{0,n}, \delta \nu_n)$ are (central frequency, $1
\over e$ width) of component $n$.  We assume that each component is an
independent physical entity and is isothermal with spin temperature
$T_{s,n}$.  This fit yields values for all $\tau_{0,n}$, $\nu_{0,n}$,
and $\delta \nu_{0,n}$. 

	Having fit the opacity profile to equation \ref{gauss1}, we next
fit the expected emission profile $T_{exp}(\nu)$ to the sum of the
emission from the CNM and WNM.  For the CNM, the brightness temperature
of the assembly of CNM components is

\begin{equation} \label{gauss2}
T_{B,CNM}(\nu) = \Sigma_0^{N-1} T_{s,n} (1 - e^{-\tau_n(\nu)}) 
	e^{-\Sigma_0^{M-1}\tau_m(\nu)} \ ,
\end{equation}

\noindent where the subscript $m$ with its associated optical depth
profile $\tau_m(\nu)$ represents each of the $M$ CNM clouds that lie in
front of cloud $n$. This fit yields values for $T_{s,n}$. For
multiple absorption components, as part of the least-squares fit process
we experiment with all possible orders along the line of sight and
choose the one that yields the smallest residuals.
 
	For the WNM, we assume that the intrinsic, unabsorbed emission
profile is represented by a set of $K$ Gaussians. We also include the
absorption of each WNM component by the CNM by assuming that a fraction
${\mathcal F}_k$ lies in front of all the CNM and is unabsorbed, with the
rest all lying behind; thus

\begin{equation} \label{gauss3}
T_{B,WNM} = \Sigma_0^{K-1} [{\mathcal F}_k + (1-{\mathcal F}_k)e^{-\tau(\nu)}]
  T_{0,k} e^{-[(\nu-\nu_{0,k})/ \delta \nu_k]^2} \ ,
\end{equation}

\noindent where the subscript $k$ represents each WNM component.  This
fit yields values for $T_{0,k}$, $\nu_{0,k}$, and $\delta \nu_{k}$. We
also experiment with determining $F_k$ from the fit as discussed in \S
\ref{cnmwnm}.  Note that $T_{0,k}$ is a brightness temperature, not a
kinetic temperature; for component $k$, the intrinsic unabsorbed
emission profile is $T_{B,WNM,k} = T_{0,k} e^{[(\nu-\nu_{0,k}/ \delta
\nu_k]^2}$.   In fitting $T_{exp}(\nu) = T_{B,CNM}(\nu) +
T_{B,WNM}(\nu)$, we usually find that ${\mathcal F}_k$ is indeterminate
and we can only distinguish between the two extremes ${\mathcal F}_k =
(0,1)$.  The differences between the ${\mathcal F}_k = 0$ and 1 profile
variances ($\sigma^2$ of the data points from the fit) are usually not
statistically significant but nevertheless lead to differences in the
derived CNM temperatures.  These differences reflect the uncertainties
in $T_{s,n}$ more than the conventional errors derived from least
squares fits. 
        
\subsection{ Permuting CNM components and trying different $\cal F$ for
WNM components} \label{cnmwnm}

	For the CNM components, $T_{B,CNM}(\nu)$ depends on the order of
the components along the line of sight.  The ordering is a discrete
quantity and cannot be solved for using the standard least-squares
technique.  Rather, we must try solutions for different orderings and
compare the variances $\sigma^2$ from the residuals of the fits.  For
$N$ CNM components there are $N!$ possible orderings.  However, some of
these are extraneous because changing the order only matters for clouds
whose profiles overlap in velocity.  Thus, we define the quantity $N_x$
such that there are $N_x!$ possible orderings of {\it overlapping} CNM
components. Obviously, $N_x \leq N$, and if $N>0$, then $N_x \geq 1$;
$N_x$ cannot be zero because a component overlaps with itself.  In our
data set, the smallest value of $N_x$ ranges from 1 to 5 (with 120
possible orderings).  The statistics are: $(N_x$, {\rm
number~of~sources}) = (1, 25), (2, 19), (3, 12), (4, 2), (5, 6); 15
sources have no detectable CNM components. 

	For the WNM components, $T_{B,WNM}$ depends on the values of
${\cal F}_k$.  ${\cal F}_k$ can, in principle, be determined by least
squares.  However, only rarely are the the variances sufficiently
sensitive to ${\cal F}_k$ for least squares to work.  Even though the
variances are insensitive, the derived parameters (such as the
associated CNM spin temperatures and the CNM and WNM HI column
densities) depend on the ${\cal F}_k$. 

	Therefore, we explore the range of each WNM's ${\cal F}_k$ by
calculating the results for three values, ${\cal F}_k = (0,0.5,1)$. 
There are $K$ WNM components and we consider all possible combinations,
which is $3^K$.  However, some of these are extraneous because the value
of a WNM component's ${\cal F}_k$ only matters if the component's
profile overlaps an absorbing CNM component.  Thus, we define the
quantity $K_x$, which is the number of such WNM Gaussian components. 
The statistics are: $(K_x$, {\rm number~of~sources}) = (1, 28), (2,
33), (3, 2), (4, 1); 15 sources have no detectable CNM components, so
$K_X=0$. 

	For any given source, the total number of trial least square
fits is $[N_x! \ 3^{K_x}]$. The two sources with the largest total number
are 3C138 and 3C409, both with $N_x=5$ and $K_x=2$ and 1080 total
trials. Most sources have many fewer trials.

	${\cal F}=0$ means that the WNM lies behind the CNM so that the
WNM'S emission is absorbed.  This means that the CNM must make up for
the missing WNM emission, so the least-square-fitted spin temperatures
are always higher for ${\cal F}=0$ than for ${\cal F}=1$.  Also, the
fitted WNM brightness tends to be larger for ${\cal F}=0$ than for
${\cal F}=1$, but this is a smaller effect.  The other WNM parameters,
central velocity and width, are only marginally sensitive to ${\cal
F}_k$.  Of course, the WNM parameters are completely insensitive to the
CNM component ordering, because we assume that ${\cal F}_k$ is the same
for all CNM components. 

\subsection{ Calculating the physical parameters}

\label{physcalc}

	For most sources, the variances don't change very much among the
various trials.  Typically these changes are less than $20\%$, and this
is not enough to distinguish one particular trial as being significantly
better than others.  (Occasionally the changes are much bigger; we
discuss these two sources below.) However, the derived physical
parameters such as spin temperature and HI column density do change
significantly among the trials.  We derive a value and error for each
physical parameter by taking a weighted average over all trials.  The
weight of each trial is the reciprocal of the variance $\sigma^2$
computed from the residuals to the fit to $T_{exp}(\nu)$. 

	For example, consider the spin temperature of a particular CNM
component for a source having $F$ trial least-square fits; normally
$F=[N_x! \ 3^{K_x}]$, but for some sources some fits don't converge and
$F$ is smaller.  Let $T_{s,f}$ be the spin temperature derived for fit
$f$ and $w_f = (1/\sigma_f^2)$ be the trial fit's weight.  Then we
derive the mean spin temperature $\langle T_s \rangle$ and its variance
$\sigma_{ \langle T_s \rangle}^2$ by the usual weighted average

\begin{mathletters} \label{wgtavg1}
\begin{equation}
\langle T_s \rangle = { \sum_{f=0}^{F-1} w_f T_{s,f} \over 
  \sum_{f=0}^{F-1} w_f  }
\end{equation}
\begin{equation} \label{varianceeqn}
\sigma_{ \langle T_s \rangle}^2 = 
\left[  { \sum_{f=0}^{F-1} w_f 
   [(T_{s,f}- \langle T_s \rangle)^2 + \sigma(T_{s,f})^2] \over 
  \sum_{f=0}^{F-1} w_f  } \right]
  \left[ {F \over F-1} \right] 
\end{equation}

\noindent Here $\sigma(T_{s,f})^2$ is the variance of $T_{s,f}$ in the
least squares fit number $f$; it accounts for the intrinsic uncertainty
in that particular fitted value for $T_{s,f}$.

	Normally one would then define the uncertainty in $\langle T_s
\rangle$ as $\Delta \langle T_s \rangle = [\sigma_{ \langle T_s
\rangle}^2 / F]^{1/2}$.  However, this is not appropriate for our case. 
This usual definition makes two assumptions: one, that the residuals
from the average, i.e.~$[T_{s,f}- \langle T_s \rangle]$, are distributed
randomly about the mean; and two, that the mean is the same for all
data.  Neither assumption is valid here.  The mean varies systematically
with ${\cal F}_k$ and tends to adopt three distinctly different values,
corresponding to the three trials of ${\cal F}_k$, with small
fluctuations about those values.  A better representation of the
uncertainty $\Delta \langle T_s \rangle$ is the spread among the three
values of $T_{s,f}$.  A good approximation to this is simply

\begin{equation} \label{simply}
\Delta \langle T_s \rangle = \sigma_{ \langle T_s \rangle} 
\end{equation}
\end{mathletters}

\noindent This description applies to all physical parameters including
HI column densities, except that we do not include the term
$\sigma(N_{f}(HI))^2$ in equation \ref{varianceeqn}. 

	Finally, we consider the most likely value $\langle {\cal F}_k
\rangle$. Again, each of the three trial values of ${\cal F}_k$ has $N_x!$
least square fits, each with variance $\sigma_f^2$. We again take a
weighted average, as above in equation \ref{wgtavg1}. However, in this
case the residuals are more nearly randomly distributed about a single
mean. Accordingly, we define the uncertainty $\Delta \langle {\cal F}_k
\rangle$ as

\begin{equation}
\Delta \langle {\cal F}_k \rangle = 
   \left[ { (\sigma_{ {\langle \cal F}_k \rangle }^2 / N_x!} \right]^{1/2} 
\end{equation}

\noindent This equation deserves a small elaboration. Consider the case
where there is no favored ${\cal F}_k$, i.e.~where the the variances for
all three choices ${\cal F}_k=(0,0.5,1)$ are equal. Then we derive 
$\langle {\cal F}_k \rangle= 0.5$,  $\Delta \langle {\cal F}_k \rangle =
0.29$. This is the maximum $\Delta \langle {\cal F}_k \rangle$; in
practice the variances don't change much with ${\cal F}_k$ so that all
sources have  $\Delta \langle {\cal F}_k \rangle$ almost this large and,
concomitantly, $\langle {\cal F}_k \rangle \sim	0.5$. 

\subsection{Limits on Opacities and Spin Temperatures for WNM
Components} \label{wnmproperties}

	In addition to parameters estimated directly from the least
squares fits, we can also place limits on the peak opacity and spin
temperature of each WNM component $k$. First, we can estimate an {\it
upper} limit to the peak opacity $\tau_{0,k}$. We do so by considering
both the intrinsic uncertainty  in the opacity profile $\Delta
e^{-\tau(\nu)}$ (equation \ref{deltatau}; downward-going solid line in
the bottom panel of the two-panel plots) and the residuals from the fit
of the CNM components to the opacity profile (solid line near
mid-profile height in the bottom panel of the two-panel plots).   These
estimates are upper limits in the sense that higher values of
$\tau_{0,k}$ would produce WNM features in the opacity profiles that are
greater than the errors in our measurements. In Table \ref{bigtable},
column 2 we list these upper limits on $\tau_{0,k}$ for WNM components.
(See \S \ref{gaussiancomps} for a description of this table.)  This
upper limit, being visually estimated, is very uncertain; its error is
not quoted, but realistically the error is large, something like half
its value. 

	The above upper limit on $\tau_{0,k}$ imply a {\it lower} limit
on $T_{s,k}$, because $T_{s,k} = {T_{0,k} \over \tau_{0,k}}$ for a WNM
component. We list these lower limits for the WNM components in Table
\ref{bigtable}, column 5 and as TSMIN in the upper panels of the
two-panel plots such as Figure \ref{zec20plot}. 

	Finally, we can estimate {\it upper} limits on on the kinetic
temperatures of Gaussian components from the fitted line widths of WNM
(and CNM) components. We define this limit as $T_{kmax}=21.86 \Delta
V^2$. This parameter is listed in Table \ref{bigtable}, column 6. For
WNM components, $T_{kmax}$ also implies a {\it lower} limit on
$\tau_{0,k}$. 

\section{GAUSSIAN COMPONENTS AND OUR DATA}

\label{gaussiancomps}

\subsection{ Gaussian parameters in tabular form and graphical form}

\label{tabular}
 
        We present the numerical results of our fits in Table
\ref{bigtable}. The opacity and expected profiles, together with other
desiderata are described in \S \ref{twopanelplots}. \begin{enumerate}

        \item Column 1: $T_B$ is the intrinsic peak brightness
temperature of the component.  For WNM components it is equal to the
unabsorbed central height of the emission Gaussian $T_{0,k}$ (see
equation \ref{gauss3}) and its error is derived as discussed in \S
\ref{physcalc}.  For CNM components, $T_B$ is equal to the spin
temperature $T_{s,n}$ times $(1- e^{-\tau_{0,n}})$ (see equations
\ref{gauss1} and \ref{gauss2}) and its error is not quoted. 

        \item Column 2: $\tau$ is the central opacity of the component. 
For CNM components, it is $\tau_{0,n}$ from equation \ref{gauss1},
derived directly from the least squares fit to the opacity profile.  
For WNM components it is the upper limit to peak opacity $\tau_{0,k}$,
estimated by eye (\S \ref{wnmproperties}), and has a very large error. 

	\item Column 3: $V_{LSR}$ is the central VLSR in km s$^{-1}$.

	\item Column 4: $\Delta V$ is the Gaussian FWHM km s$^{-1}$.

	\item Column 5: $T_s$ is the spin temperature. For CNM
components it is derived from the fit to the expected profile; its error
is derived as discussed in \S \ref{physcalc}. For a few CNM components,
$T_s$ had to be forcibly set to zero to attain convergence of the fit to
the expected profile.  For WNM components $T_s$ is a lower limit (\S
\ref{wnmproperties}), and has a very large error; this is TSMIN in the
two-panel plots. 

	\item Column 6: $T_{kmax}$ is the kinetic temperature of a
component if there were no nonthermal broadening; $T_{kmax} = 21.86 \Delta
V^2$.
        
	\item Column 7: $N(HI)_{20}$ is the HI column density in units
of $10^{20}$ cm$^{-2}$. For the few CNM components with $T_s=0$, we
assign $N(HI)=0$.

	\item Column 8: For WNM components. ${\cal F}$ is the weighted
average of ${\cal F}$, calculated as described in \S \ref{physcalc}. For
CNM components, O is equal to the order of the component along the line
of sight, beginning with 0; increasing numbers mean increasing distance
along the line of sight. This order is the one for which the variance
from the fit to the emission spectrum is smallest. For CNM components,
no error is given.
        
	\item Column 9: $l/b/$SOURCE are the Galactic longitude,
latitude, and source name.
        
	\end{enumerate}

\subsection{ The process of fitting Gaussians}

\label{gaussianprocess}

	First, we remark that the astronomically popular activity of
fitting Gaussians to spectral profiles is based on the assumption that
velocity profiles are, indeed, characterized by Gaussians. This is
necessarily true when the damping wings are insignificant, as for the
21-cm line, and when there are no nonthermal motions. However, most of
our Gaussian components have linewidths $T_{kmax} > T_s$, so that the
line shape is determined by nonthermal motions. Unless the nonthermal
motions consist of a large number of turbulent elements, so that the
central limit theorem applies, there is no known reason why a
component's shape need be Gaussian. Nevertheless, both the opacity and
expected profile of most sources are easily decomposed into Gaussians,
so whether or not this model is physically correct it works empirically
and is convenient---and besides, how else would one proceed?

	Gaussian functions are not orthogonal, so our Gaussian fits are
not unique.  Rather, they reflect our subjective judgment in selecting
the number of Gaussian components and other biases.  Here we describe
our approach to this subjective procedure, a procedure based on the
equations of \S \ref{gaussrad}. 

	For each source we begin by fitting Gaussians to the opacity
spectrum (equation \ref{gauss1}; we call these the ``CNM Gaussian
components''.  Our approach is to use the minimum number of components
required to make the residuals of the fit comparable to or smaller than
the errors derived as discussed in \S\ref{realerrors}.   After fitting
the opacity components, we fit the expected profile by simultaneously (1)
fitting the intensities of the opacity components (whose
centers and widths are fixed) and (2) adding one or more ``WNM Gaussian
components'', which contribute to the expected profile but not to the
opacity profile. This fit to the expected profile is a fit to the sum
of equations \ref{gauss2} and \ref{gauss3}.

	For the WNM we again use the minimum number of WNM components.
However, in this case the criterion involves reducing the residuals of
the fit to a ``reasonable'' level instead of to the much smaller level
of the derived errors. The reason is that the WNM components are derived
from the observed expected profile, and this  is subject to an
unaccounted-for source of error, namely the contribution of HI emission
from directions outside the main telescope beam. In fact, almost half of
the response of the Arecibo telescope to an extended source, such as the
21-cm line, comes from outside the main beam (Heiles et al 2001a). These
directions are likely to contain velocity components that are unrelated
to those within the main beam, or have Gaussian parameters that are
modified by angular derivatives. 

	Consequently, when fitting the expected profile we use our
subjective judgment in deciding how many WNM components to fit. Almost
always this number is small, consisting of one or two fairly narrow WNM
components often sitting on top of a much weaker, broader one. Below we
present and discuss a number examples to illustrate our approach and our
subjective definition of what is ``reasonable''.

	During the fitting process we were concerned about ``stray
radiation'', which is radiation entering the telescope from distant
sidelobes. This can produce unreal features in observed emission
spectra. These are usually broad and could cause us to fit weak, broad
WNM Gaussians that are unrelated to the gas in direction observed. For
each source we compared the tails of our expected profiles
$T_{exp}(\nu)$ with the nearest profile of the Leiden-Dwingeloo survey
(Hartmann and Burton 1997). We were gratified---and somewhat surprised
in view of Arecibo's substantial aperture blockage (Heiles et al
2001a)---to find that the differences between the intensities in the
profile wings for the two datasets are completely negligible. This
implies that Arecibo's significant sidelobes all lie close to the
direction of the main beam. Thus, our HI column densities, even for the
weak, broad WNM components, are quite reliable. 

\subsection{Optical lines versus the 21-cm line}

\begin{figure}[h!]
\begin{center}
\includegraphics[width=3.5in]{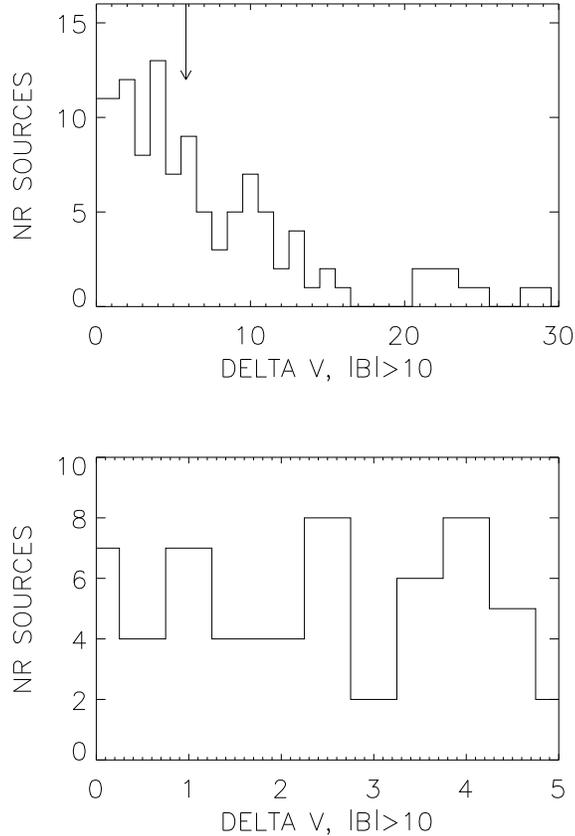} 
\end{center}

\caption{Histogram of the velocity separations between adjacent CNM
Gaussians. The lower panel is an expanded version. \label{hist_deltav}} 
\end{figure}

	Our results should be comparable to the results from optical
absorption lines of minority ionization species such as KI because both
sample the CNM. The 21-cm line opacity $\propto N(HI)
T_k^{-1}$. The KI opacity is approximately $\propto N(KI)$; under
ionization equilibrium, $N(KI) \propto N(HI) P T^{-1.7}$, where $P$ is the
CNM pressure, which is typically $(P/k) = 2250$ cm$^{-3}$ K (Jenkins \&
Tripp 2001). Thus optical absorption lines are biased towards low
temperatures even more than the 21-cm line is.

	Optical interstellar absorption lines sample heavy elements,
which have smaller thermal motions than HI, so they might reveal
multiple narrow components where we might resolve only one. If we take
the results at face value, then this is indeed occurring. Figure
\ref{hist_deltav} exhibits the histogram of velocity separations between
adjacent CNM Gaussian components. Welty and Hobbs (2001; WH) present the
same histogram for KI optical absorption line components in their Figure
16. Their histogram has a pronounced peak centered at about 1.8 km
s$^{-1}$, falling to zero below 0.7 km s$^{-1}$ and  becoming small
above 3 km s$^{-1}$. In contrast, ours is roughly flat from 0 to 5 km
s$^{-1}$. They discuss their histogram in terms of a continuous
distribution of closely-spaced components, and the difference between
our histogram and theirs could be a result of blending. 

	However, we have lingering doubts about the reality of some of
their closely-spaced components. In some cases they need several
components to reduce the residuals to their noise level; in other cases
they do not. To us, their sample of closely-spaced components seems
inconsistently nonuniform. Looking at their Table 1, many stars have
closely-spaced components and many do not  (defined as $< 1.9$ km
s$^{-1}$ for this discussion). For example, $\omega^1$Sco at $(l,b) =
(352.8^\circ, 22.8^\circ)$ has 4 while the fairly nearby 1Sco at
$(346.1^\circ, 21.8^\circ)$ has none; the two stars have comparable
visual line profiles and extinctions $E(B-V)= (0.22, 0.19)$ mag,
respectively. Are conditions really that different towards the two
stars, or is the presence of multiple closely-spaced components a vagary
of small departures from idealized Gaussian line shapes and subjective
judgment? 

	We asked this question above about our own Gaussians in \S
\ref{gaussianprocess} and the question applies equally to the optical
data. The KI line profiles  are always dominated by nonthermal motions,
because even the multiple closely-spaced components have nonthermal line
widths. WH prefer to decompose  the line shapes by inserting multiple
Gaussians subcomponents (actually, Voigt subcomponents) until the
residuals are equal to the noise. This is a valid mathematical
description of the profile, although it is not unique---but is it a
valid {\it physical} description? We repeat, there is no physical reason
to assume that turbulent broadened profiles are, in fact, Gaussian,
unless the number of subcomponents is so large that the central limit
theorem applies.

	In our opinion, the use of many subcomponents to represent a
profile peak might, but does not necessarily, describe the physical
world. 

\subsection{Errors incurred from fitting too many/too few Gaussians}

	The nonuniqueness of Gaussian fitting means that we might
sometimes fit opacity profiles with too many or too few Gaussians. For
example, perhaps the multiple closely-spaced components of WH are
physically real. Here we discuss how this affects our derived spin
temperature $T_s$.

\subsubsection{The low-$\tau(\nu)$ case}

	We assume $\tau = (1 - e^{\tau)}$, which is valid for $\tau
(\nu) \lesssim 1$. Whether we fit $N$ overlapping Gaussians, or fit only
one when $N$ are required in the physical world, we always require
agreement with the data, i.e.

\begin{mathletters}
\begin{equation} \label{tausum}
\sum_0^{N-1} \tau(\nu)_n = \tau(\nu)
\end{equation}
\begin{equation} \label{tbsum}
\sum_0^{N-1} T_{s,n} \tau(\nu)_n = T_{exp}(\nu)
\end{equation}
\end{mathletters}

\noindent where $n$ represents each Gaussian and the quantities on the
right are the observed ones. The easiest way to satisfy this combined
requirement on both sums is for $T_{s,n}$ to be the same for all
components, and this makes $T_s$ independent of $N$. If $T_{s,n}$
changes with $n$, then the values should cluster around the
$n$-independent one. Therefore, derived spin temperatures for all the
Gaussian components are not sensitive to the number of fitted Gaussians
$N$.

	The same cannot be said of the velocity widths and integrated
areas of the Gaussians. If we fit Gaussian-shaped opacity and expected
profiles with multiple blended Gaussians, then the widths ($T_{kmax}$) and
areas ($N(HI)$) necessarily decrease with $N$.

\subsubsection{The high-$\tau(\nu)$ case}

	The high-$\tau(\nu)$ case is harder to treat because the
emission is nonlinearly related to $\tau$. Here we consider a simple
model in which $N$ clouds of identical $\tau(\nu)$ and $T_s$ lie behind each
other. In this case, equation \ref{tausum} remains valid but equation
\ref{tbsum} becomes a simpler version of equation \ref{gauss2} with all
$T_{s,n}=T_s$ and all $\tau(\nu)_n = \tau$, namely

\begin{equation}
T_{exp}(\nu) = T_s \Sigma_0^{N-1}(1 - e^{-\tau(\nu)}) 
	e^{-\Sigma_0^{M-1}\tau(\nu)} \ ,
\end{equation}

\noindent One can manipulate this equation to show that $T_s$ is
independent of $N$, as it is for the low-opacity case. Thus again, for
this illustrative model the derived spin temperature is unaffected by
the number of components that represent the profile.

\begin{figure}[p!]
\begin{center}
\includegraphics[height=6in] {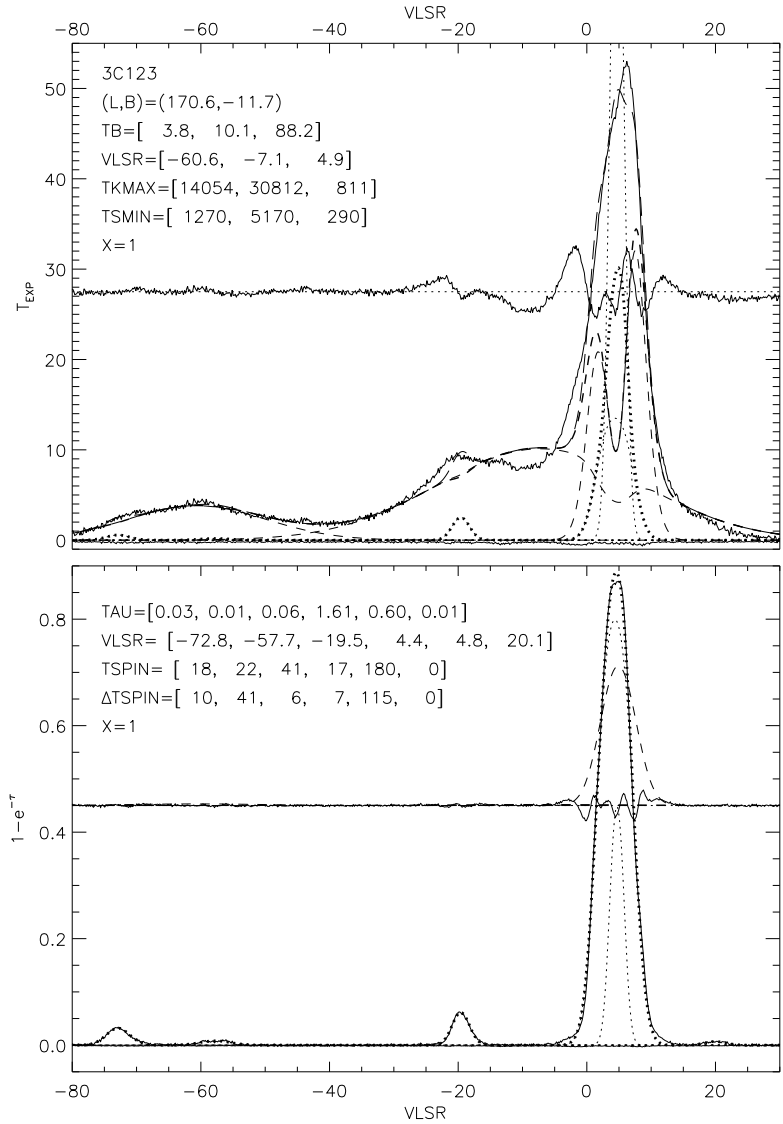} 
\end{center}

\caption{Data for 3C123.  Profile uncertainties are invisible because
the source is strong. For a complete description of this figure, see
Figure \ref{zec20plot}, \S \ref{twopanelplots}, and \S \ref{tabular}. 
\label{3C123twopanelxx}} \end{figure}

\begin{figure}[p!]
\begin{center}
\includegraphics[height=6in] {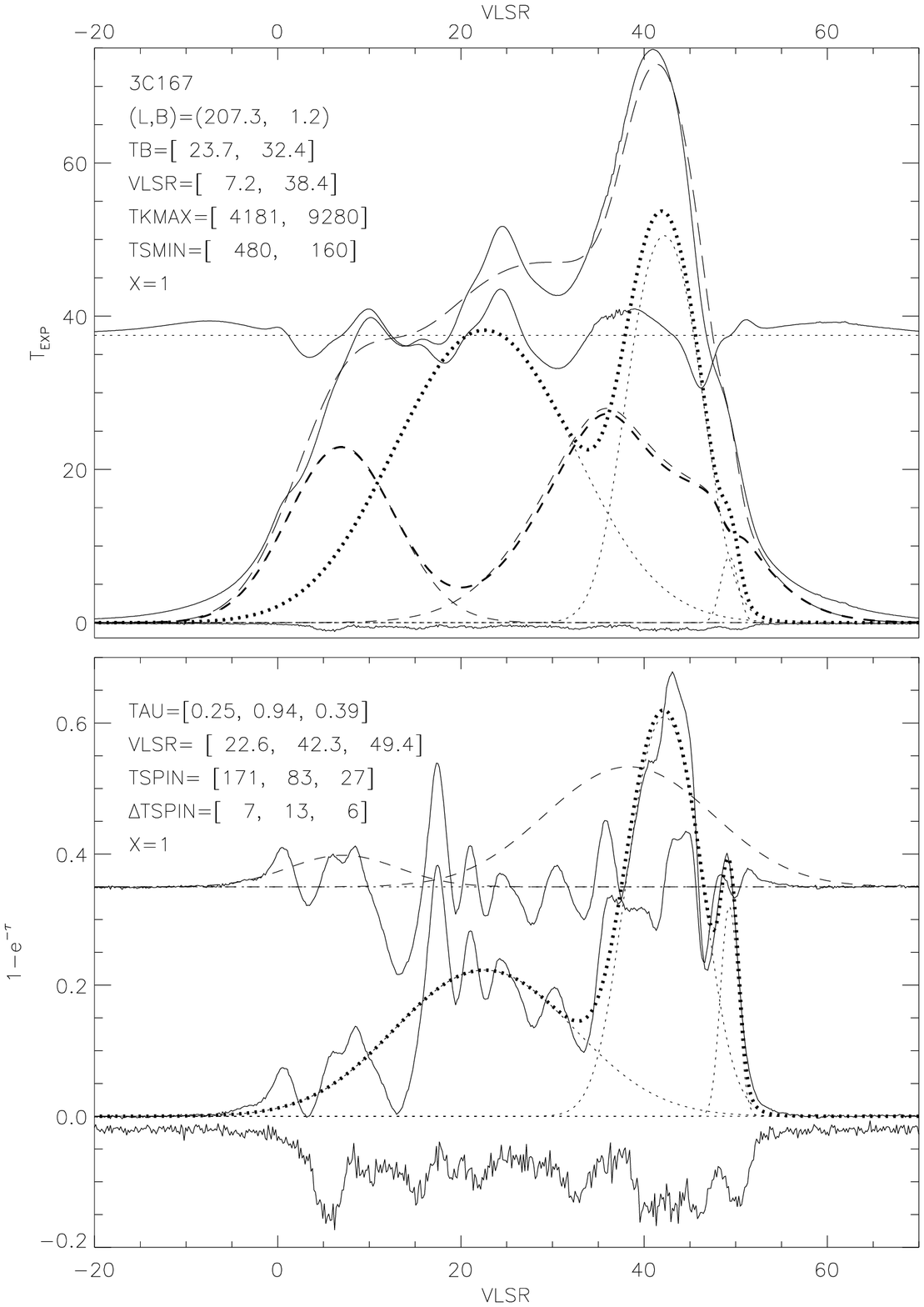} 
\end{center}

\caption{Data for 3C167.  For a complete description of this figure, see
Figure \ref{zec20plot}, \S \ref{twopanelplots}, and \S \ref{tabular}. 
\label{3C167twopanelxx}} \end{figure}

\begin{figure}[p!]
\begin{center}
\includegraphics[height=6in] {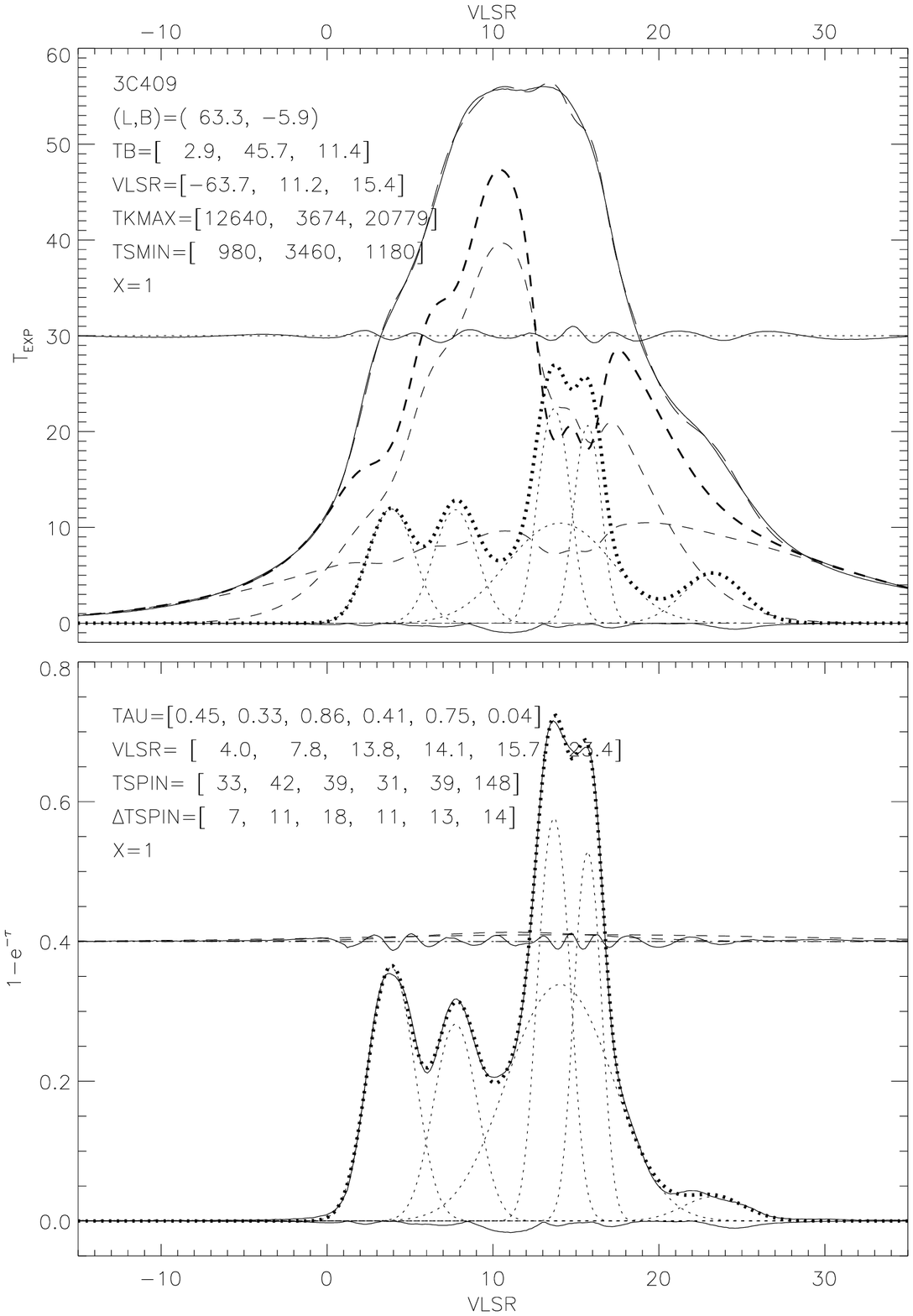} 
\end{center}

\caption{Data for 3C409.  For a complete description of this figure, see
Figure \ref{zec20plot}, \S \ref{twopanelplots}, and \S \ref{tabular}. 
\label{3C409twopanelxx}} \end{figure}

\begin{figure}[p!]
\begin{center}
\includegraphics[height=6in] {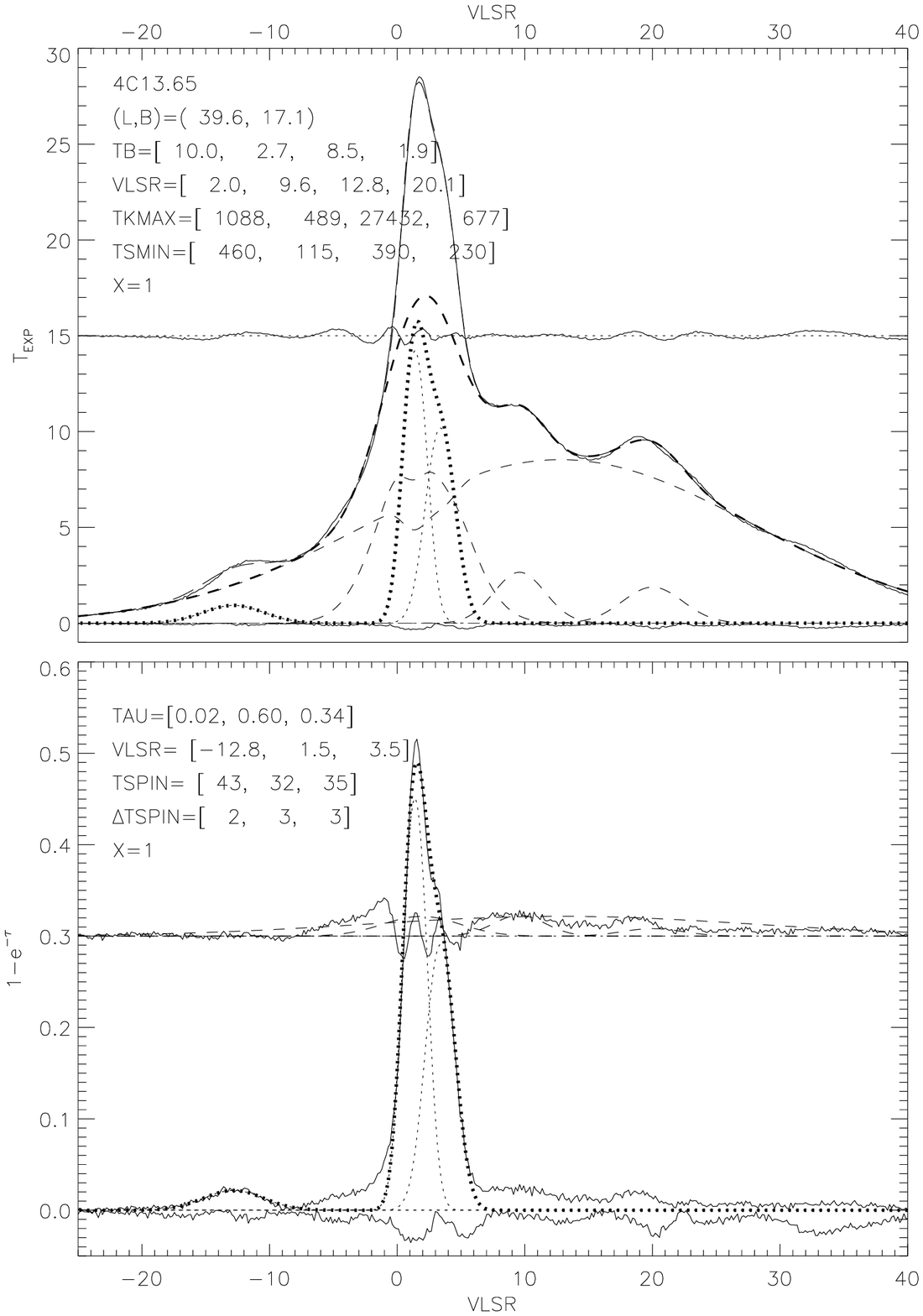} 
\end{center}

\caption{Data for 4C13.65.  For a complete description of this figure, see
Figure \ref{zec20plot}, \S \ref{twopanelplots}, and \S \ref{tabular}. 
\label{4C13.65twopanelxx}} \end{figure}

\begin{figure}[p!]
\begin{center}
\includegraphics[height=6in] {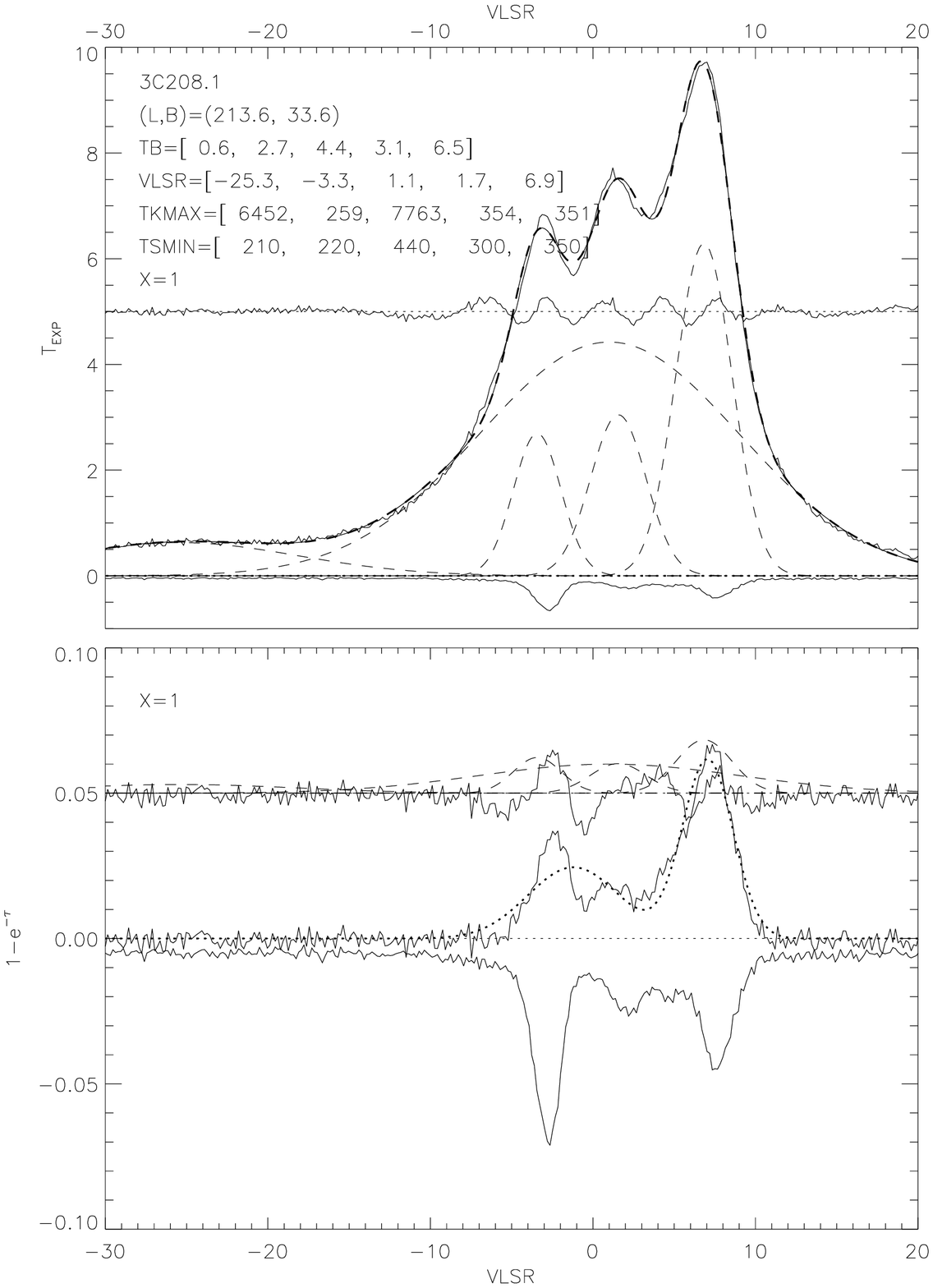} 
\end{center}

\caption{Data for 3C208.1.  For a complete description of this figure, see
Figure \ref{zec20plot}, \S \ref{twopanelplots}, and \S \ref{tabular}. 
\label{3C208.1twopanelxx}} \end{figure}

\begin{figure}[p!]
\begin{center}
\includegraphics[height=6in] {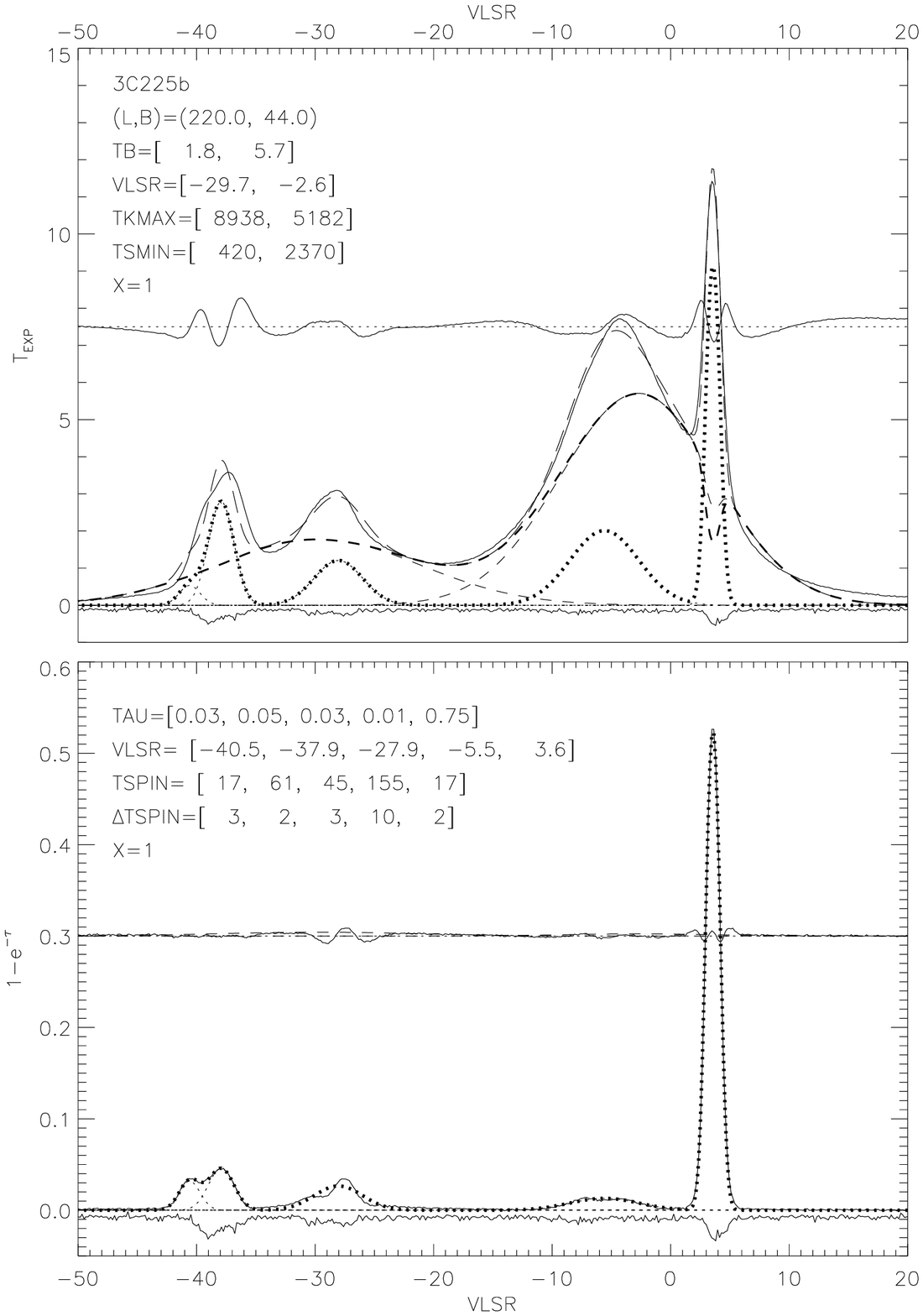} 
\end{center}

\caption{Data for 3C225b.  For a complete description of this figure,
see Figure \ref{zec20plot}, \S \ref{twopanelplots}, and \S
\ref{tabular}.  \label{3C225btwopanelxx}} \end{figure}

\begin{figure}[p!]
\begin{center}
\includegraphics[height=6in] {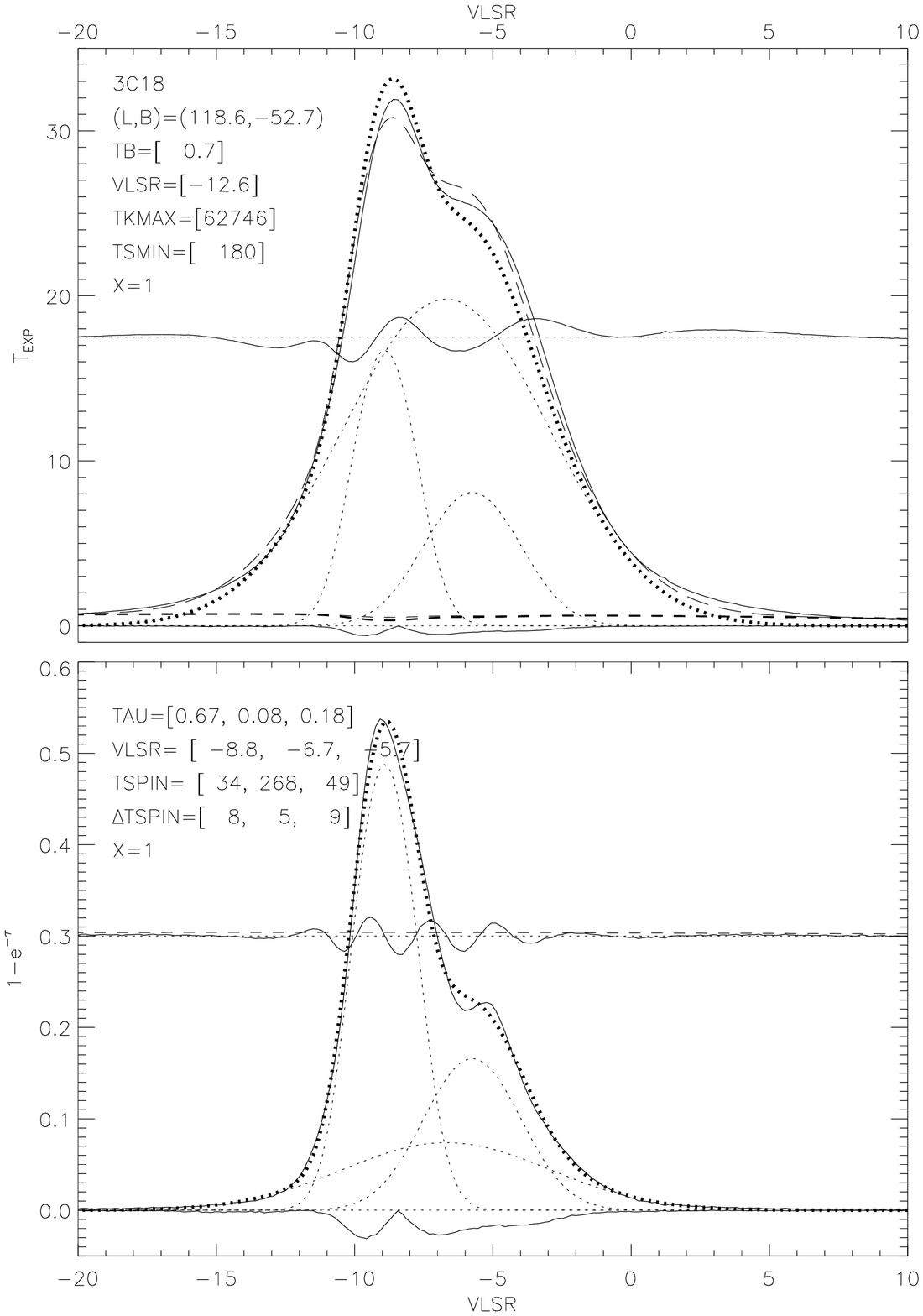} 
\end{center}

\caption{Data for 3C18.  For a complete description of this figure,
see Figure \ref{zec20plot}, \S \ref{twopanelplots}, and \S
\ref{tabular}.  \label{3C18twopanelxx}} \end{figure}

\begin{figure}[p!]
\begin{center}
\includegraphics[height=6in] {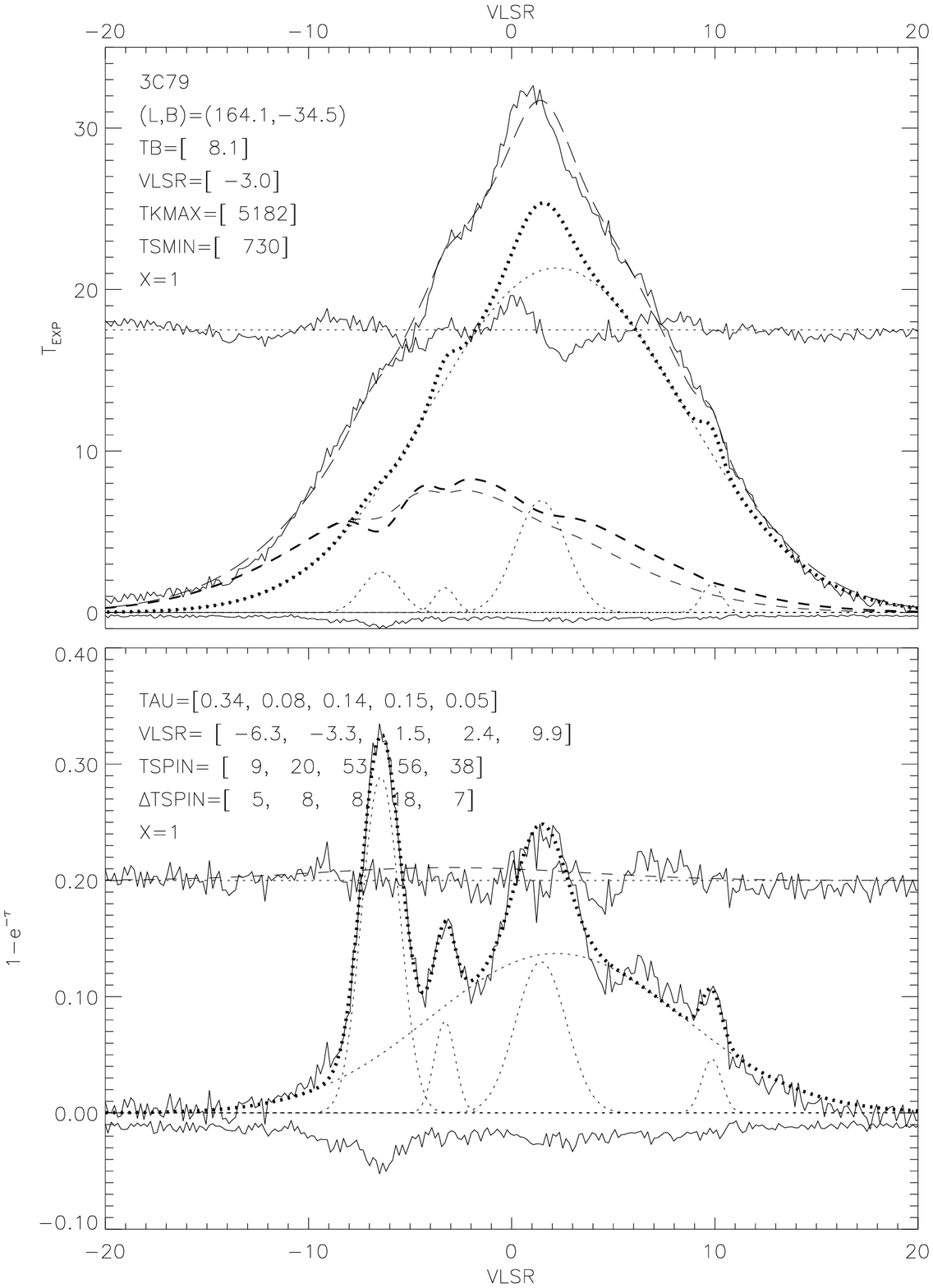} 
\end{center}

\caption{Data for 3C79.  For a complete description of this figure,
see Figure \ref{zec20plot}, \S \ref{twopanelplots}, and \S
\ref{tabular}.  \label{3C79twopanelxx}} \end{figure}

\begin{figure}[p!]
\begin{center}
\includegraphics[height=6in] {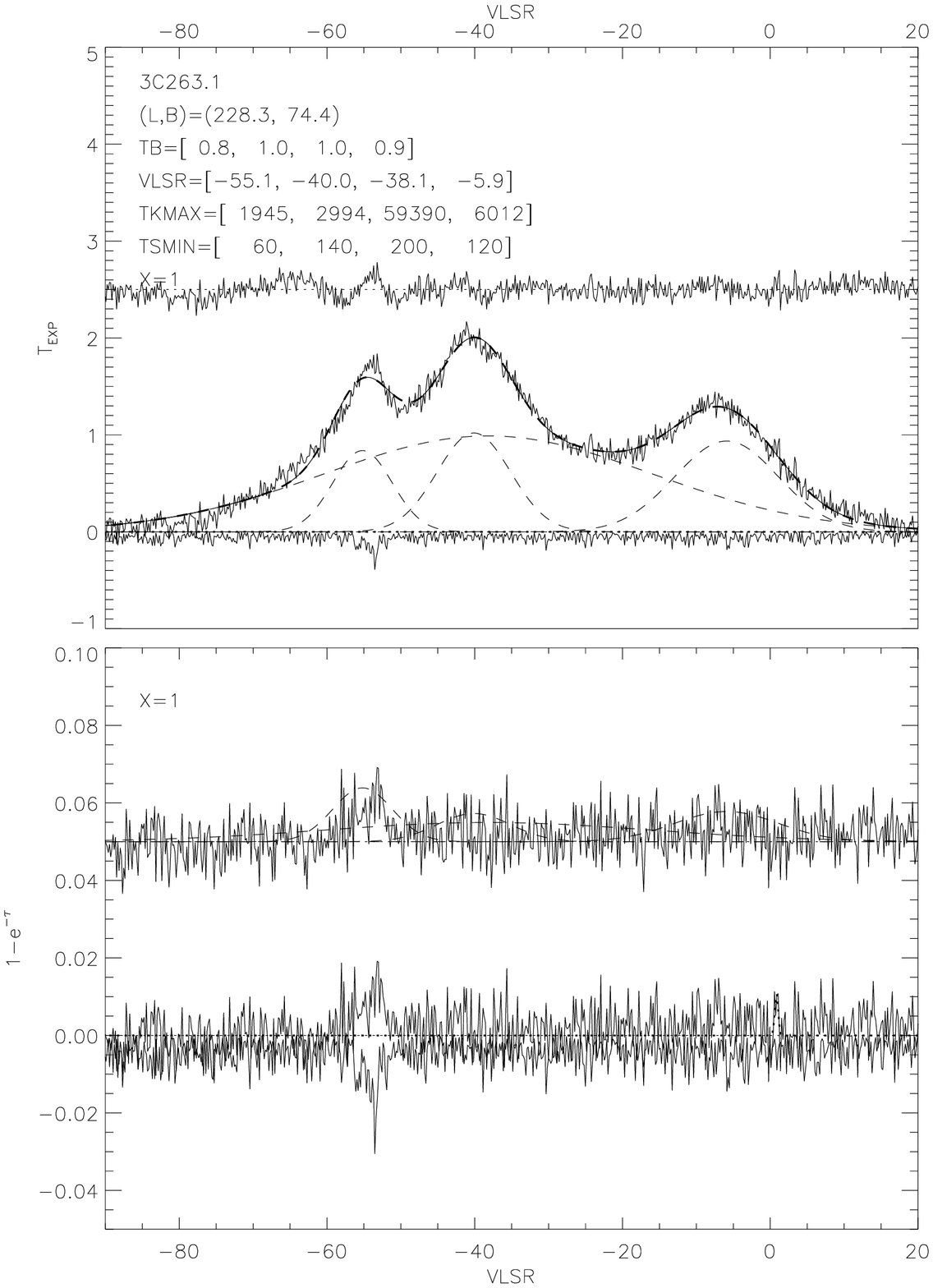} 
\end{center}

\caption{Data for 3C263.1.  For a complete description of this figure,
see Figure \ref{zec20plot}, \S \ref{twopanelplots}, and \S
\ref{tabular}.  \label{3C263.1twopanel00}} \end{figure}

\begin{figure}[p!]
\begin{center}
\includegraphics[height=6in] {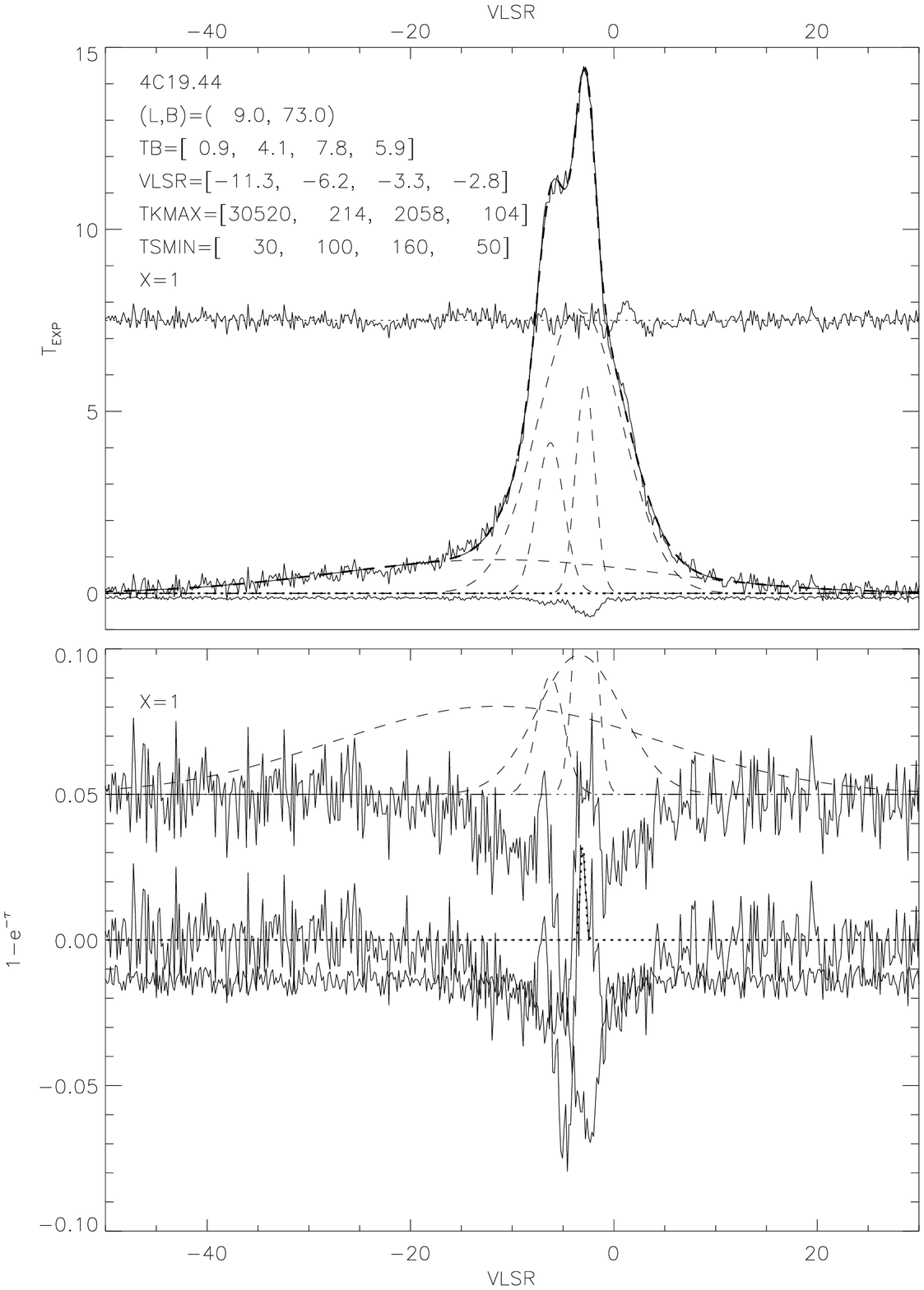} 
\end{center}

\caption{Data for 4C19.44.  For a complete description of this figure,
see Figure \ref{zec20plot}, \S \ref{twopanelplots}, and \S
\ref{tabular}.  \label{4C19.44twopanel00}} \end{figure}

\begin{figure}[p!]
\begin{center}
\includegraphics[height=6in] {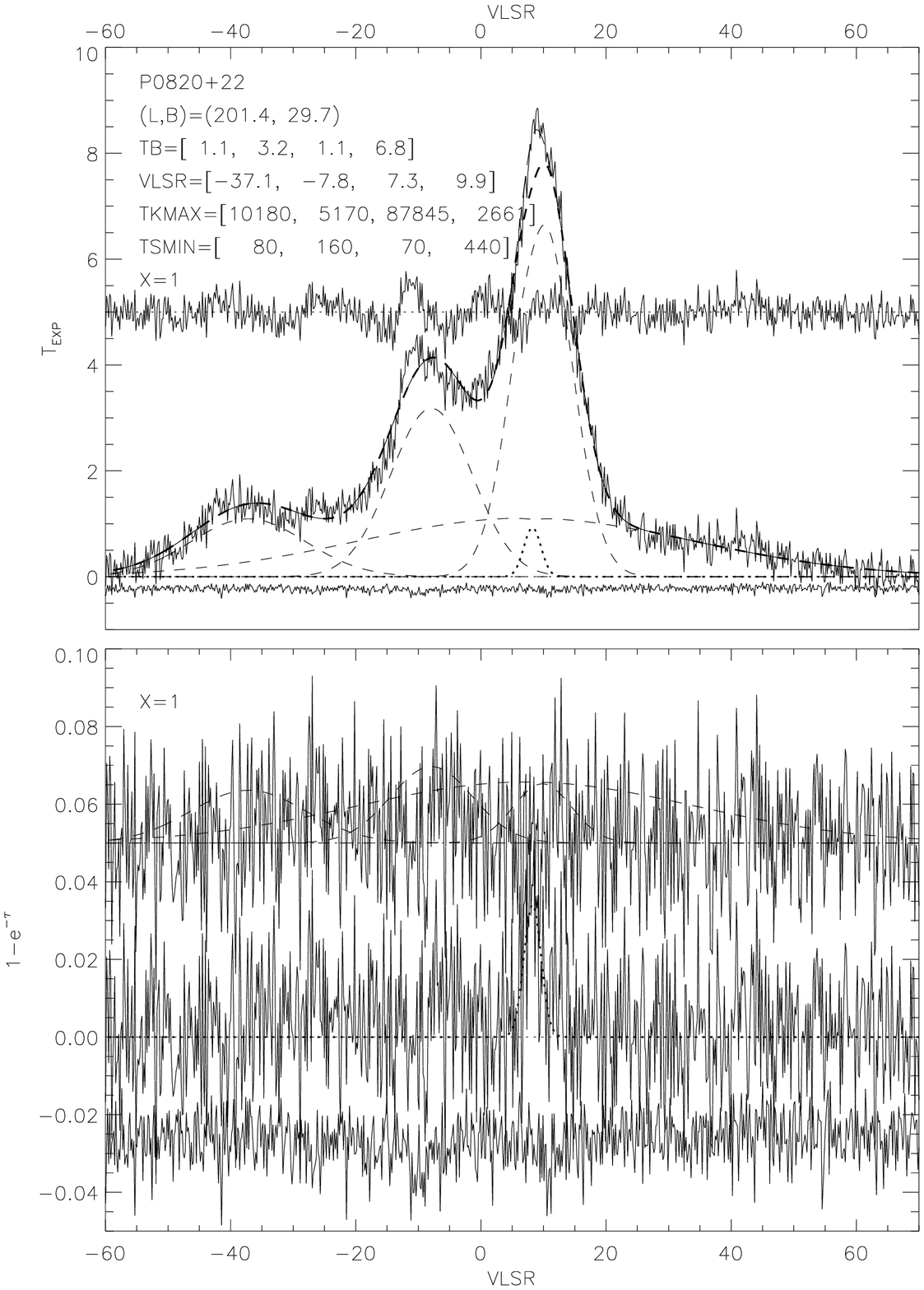} 
\end{center}

\caption{Data for P0820+22.  For a complete description of this figure,
see Figure \ref{zec20plot}, \S \ref{twopanelplots}, and \S
\ref{tabular}.  \label{P0820+22twopanel00}} \end{figure}

\subsection{Fitting the opacity profiles: CNM Gaussian components}

\subsubsection{Sources for which the CNM fits are fairly unambiguous}

	For many sources either a single Gaussian or multiple ones that
don't overlap are sufficient (i.e., $N_x=1$); examples are 3C315 and
3C207, shown in Figures \ref{3C315abplot} and \ref{3C207pubplot} below.
These cases are unambiguous and there is very little room for mistakes,
except for the possibility that each Gaussian really consists of more
than one narrower and/or weaker ones.  26 out of the 62 sources having
CNM components fall into this unambiguous category. 

	Many sources have overlapping Gaussians. Examples are 3C120 and
P0531+19, shown in Figures \ref{3C120abplot} and \ref{P0531+19_pubplot}
(these are presented below because their main focus is in later
discussion).  For 3C120 we use three Gaussians and achieve residuals
that are considerably smaller than the errors; this exceeds the
requirement of a good fit.  However, two Gaussians produce residuals
that vastly exceed the errors, so three Gaussians is the correct choice.
 We have not been able to fit three Gaussians with other parameters than
those given, so the validity of this fit seems secure.  P0531+19 has two
distinctly separated narrow peaks connected by a nonzero broad region
and is fit well by two narrow and one broad Gaussians. Three Gaussians
are certainly required and, again, no other combination of Gaussians
works, so this seems the correct solution. 

\subsubsection {Sources for which the CNM fits are more ambiguous or
difficult} 

3C123 (Figure \ref{3C123twopanelxx})  is a good example of two
difficulties. One is overlapping opacity components, which make the
uniqueness of a fit less certain; in the case of 3C123, the overlapping
components are indistinguishable, making a true fit impossible.  3C123
is a very strong source, so its opacity spectrum has small errors and
imperfections of the fit reveal small inadequacies. The main opacity
peak has ($\tau_0= 1.60 \pm 0.03$) and, when fit with a single or even a
``reasonable number'' of Gaussians, has residuals that vastly exceed the
errors. We believe that the main absorption peak consists of several
overlapping narrow components, which are degenerate in the fitting
process.  3C123 lies near a molecular cloud in the Taurus/Perseus region
and even exhibits OH absorption, which is rare for our sources. 3C123
has at least three narrow OH absorption components ($\Delta V = 0.5$ to
1.4 km s$^{-1}$) in the velocity range 2 to 7 km s$^{-1}$ (Troland \&
Heiles, in preparation; Crutcher, Troland, \& Heiles 1981). Each of
these is likely to have associated HI, producing overlapping narrow HI
components. 

	An additional problem with 3C123 is its 20.0 km s$^{-1}$ CNM
component: the least squares fit to the expected profile would not
converge unless its spin temperature was forced to zero. Its opacity
$\tau_0=0.007 \pm 0.002$, which is very small, and we suspect that this
component occupies a small solid angle, absorbing a small fraction of
3C123 continuum and contributing insufficient 21-cm line brightness to
be detected. The same situation occurs for two opacity components in 3C273.

	Sources near the Galactic plane (defined here as $|b|<10^\circ$)
have complicated absorption spectra and cannot always be fit in a
reasonable way; 3C167 (Figure \ref{3C167twopanelxx}) is an example,
others being 3C141.0, T0526+24, T0556+19, T0629+10). On the other hand,
some low-$|b|$ sources are well-fit: for example, 3C409 (Figure
\ref{3C409twopanelxx}) is exceedingly well-fit, seemingly unambiguously,
by 6 Gaussians; others (not quite so well fit) include 3C154, 4C13.67,
4C22.12, P0531+19, 3C138. Intermediate cases are 3C131, 3C133, 3C410. 

\subsection{Fitting the expected profiles: WNM Gaussian components}

	We thoroughly discuss two example cases, 3C120 and 3C138, in
detail in \S \ref{unusualsources}. Here we present some examples and
describe them only briefly.

\subsubsection{ Sources for which WNM fits are fairly unambiguous}

	Sometimes  only a single WNM component is required (e.g.\
P0531+19, Figure \ref{P0531+19_pubplot}).  More often two are required,
one a broad, weak one to fit the wings of the line and the other a much
more intense and narrow component to fill in the emission line, e.g.\
3C207 and 3C315 (Figures \ref{3C207pubplot}, \ref{3C315abplot}).  
Sometimes additional nonoverlapping WNM components at different
velocities are required, e.g.\ 3C120 (Figure \ref{3C120abplot}). All
these cases are straightforward and unambiguous.  More complicated WNM
cases include, of course, sources near the Galactic plane, and also
others. Many complicated cases are nevertheless well-fit, including
those listed above as being unambiguously fit with CNM Gaussians. 

\subsubsection {Sources for which the WNM fits are more ambiguous or
difficult} 

	Some sources away from the plane are remarkably complicated:
4C13.65 (Figure \ref{4C13.65twopanelxx}) has a broad WNM component with
several narrower ones within; it is complicated but the Gaussian process
seems fairly secure.  3C208.1 (Figure \ref{3C208.1twopanelxx}) has no
reliable CNM components because of the relatively large $\Delta
e^{-\tau(\nu)}$, but it has several WNM components; its nearby brother,
3C208.0, is similar. 

	3C225a (Figure \ref{zec20plot}) and 3C225b (Figure
\ref{3C225btwopanelxx}) have truly complicated expected profiles, each
of which extends over a wide velocity range and has several peaks
connected by a broad Gaussian which is only a surrogate for reality.
These sources are separated by only 6.3 arcmin so that one expects  them
to be similar, and they are. However, their Gaussian representations
differ significantly, partly because 3C225b has CNM components at
velocities where 3C225a has WNM ones; this is, in turn, a consequence of
the fact that 3C225a has large errors in its opacity spectrum (Figure
\ref{zec20plot}). Also, the fitting process yields quite different
estimates for the WNM components in the velocity range --15 to 10 km
s$^{-1}$, largely because the contribution to the expected profiles of
the broad weak CNM component in this range is poorly determined,
particularly for 3C225a. Inconsistencies between the fits to this pair
of sources illustrate the uncertainties introduced by large angular
derivatives. Fortunately, our observing and analysis techniques reveal
these derivatives and the errors they introduce. Because the optical
depth error profile for 3C225b is about a factor of two smaller than
that for 3C225a, the  WNM and CNM components to 3C225b should be more
reliable. 

	Heiles (2001b) presented a short preliminary report of this work
which featured 3C18 (Figure \ref{3C18twopanelxx}) as an example of a
straightforward fit.  Ironically, further reflection reveals that it is
not so straightforward, despite the fact that its profiles look so
simple; in fact, 3C18 is probably the most difficult case in our entire
source list with regard to the ambiguity in assigning CNM and WNM
components. To illustrate this ambiguity for such a difficult case we
discuss the alternative fit in the following paragraph. 

	In the  current fit for 3C18 we have replaced the earlier narrow
WNM component, which had $T_{kmax} = 2200$ K, with a CNM component that
is $30\%$ wider with $T_{kmax} = 3640$ K and $T_s = 586 \pm 11$ K.  With
the earlier fit, the residual of the opacity profile fit (i.e., the
difference between the observed and fitted opacity profiles) had the
same shape as the original narrow WNM component and was somewhat larger
in magnitude than the  $\Delta \tau(\nu)$ profile. Accordingly, we could
assign the residual to a new CNM component. At the same time, we
eliminated a doubtful narrow CNM component in the original fit.  The
original and new fits are of comparable quality. This illustrate the
effects of subjective judgment on measured spectra whose fitted
residuals lie close to the intrinsic errors.  We emphasize that we do
not necessarily regard the current fit as better than the earlier one;
rather, we present this difficult case as an illustration. Accordingly,
the spin temperature of this CNM component should be regarded with
suspicion. However, if its spin temperature is close to the derived
value of $267 \pm 5$ K, then this source is very unusual in having
insignificant WNM.  Our opinion is that the spin temperature has little
validity, because all other sources have significant WNM column density
fractions.  This change doesn't affect any other parameters much, nor
does it affect the conclusion in Paper II that $T_{kmax}$ for this
component lies in the thermally unstable range. 

   	The fitting of Gaussians to some sources seems strained.  3C79
(Figure \ref{3C79twopanelxx}) is one example with a bumpy opacity
profile and a nearly triangular expected profile. We fit it
successfully with 5 CNM components, which seems excessive for a source
at Galactic latitude $b= -35^\circ$, and a single WNM component.
      
\subsubsection{ Sources with no CNM components}

	19 sources exhibit no detectable opacity profile; we fit them
only with WNM components. Most of these fits are straightforward,
requiring one or two rather clearly-defined components. Some are more
complicated. 3C263.1 (Figure \ref{3C263.1twopanel00}) has three narrower
WNM components embedded in a much broader one; the need for the three is
clear, but the broad one might be an unreal way to represent more
blended components.  P0820+22 (Figure \ref{P0820+22twopanel00}) is
similar, but it also contains one or two WNM components that are
considerably narrower, with widths comparable to typical CNM components.
4C19.44 (Figure \ref{4C19.44twopanel00}) has three narrow blended WNM
components; again, their widths are similar to those of typical CNM
components, but the errors in the opacity spectrum prevent them from
being detected.

\subsection{ Two illustrative examples of the combined Gaussian fitting
process} \label{unusualsources}

	Here we discuss the fits for two particular sources in some
detail to illustrate our Gaussian-fitting process. We choose 3C120 and
3C315 because they are typical in most respects, but are unusual in the
following way. For most sources the variances do not change much for
different trials of ${\cal F}_k$; the ratios of maximum to minimum
variance are usually less than 1.2. However, some sources have
significantly larger ratios, and these two sources have the largest, 3.1
and 2.9 respectively. The next highest is 4C19.44 (ratio=1.8). 

	First a comment on sources that have atypical ratios that exceed
1.4: 3C120, 3C138, 3C315, 3C318, 3C353, 3C93.1, 4C19.44, NRAO140. Some
sources are so extreme that their fits don't even converge for some
trials; this happened for 3C123, 3C138, 3C172.0, 3C315, NRAO140. For
most of these sources, the fits that do converge produce variances that
don't differ much. 

	We display the two extreme fits for 3C120 and 3C315 in plots
with three panels in Figures \ref{3C120abplot} and \ref{3C315abplot}. In
the top two panels of these figures we plot the observed expected
profile $T_{exp}(\nu)$ and the two extreme fits.  For both sources the
residuals are smaller when the principal WNM component has ${\cal F}=0$,
i.e.\ it lies fully behind the absorbing CNM components. The absorption
of the broad WNM component by the narrower CNM ones puts more structure
on the central peak of the profile. 	

\subsubsection{ 3C120}

\begin{figure}[p!] 
\begin{center}
\includegraphics[height=6in] {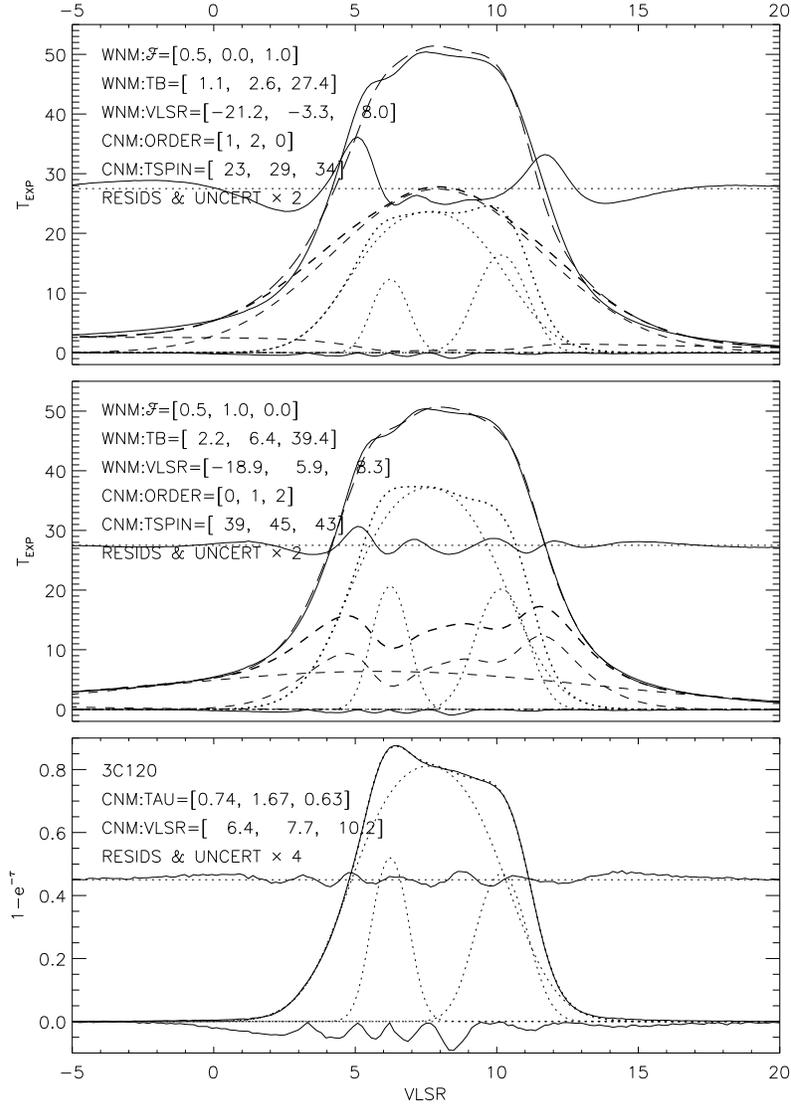} 

\caption{Data for 3C120.  Top two panels: data and fits having largest
(top) and smallest residuals for the trial fits to $T_{exp}(\nu)$. 
Bottom panel: data and fits for the opacity spectrum.  For a more
detailed description of this figure, see \S\ref{unusualsources}.  Here,
the top two panels are similar to the top panel in Figure
\ref{zec20plot}, and the bottom panel is like the bottom panel in that
figure. \label{3C120abplot} }

\end{center}
\end{figure}

	Figure \ref{3C120abplot} exhibits the results for 3C120.  Note
the factors of two and four expansion in vertical scale for the
residuals and uncertainties on the top two and bottom panels,
respectively.  First, some general comments about the choice of Gaussian
components.  The opacity spectrum in the bottom panel is fit very well
by the three blended CNM components, with residuals smaller than the
errors; fitting only two CNM Gaussians makes the residuals unacceptably
large.  The emission spectrum is fit by three WNM Gaussians, one of
which is weak and centered at --19 km s$^{-1}$, outside the $V_{LSR}$
plotting range.  The other two WNM components are blended: one is wide
and weak, providing a floor over which the narrower, more intense WNM
Gaussian sits. 

	The residuals and uncertainties are expanded by a factor of two
on the top two panels.  The differing residuals are obvious.  The
residuals are affected almost exclusively by ${\cal F}$; the ordering of
CNM components affects the variance only at the $\sim 2\%$ level,
despite the large CNM central optical depths (1.67, 0.74, 0.63).  The
relative insensitivity of the residuals to CNM ordering is typical. 

	The intense, narrow WNM Gaussian has halfwidth $6.75 \pm 0.78$
km s$^{-1}$, corresponding to $T_{kmax} = 1000$ K.  From Table
\ref{bigtable}, its peak intrinsic weighted mean $T_{0,k} = 41.3 \pm
5.6$ K, so that its peak opacity $\tau_0 \geq (41/1000)= 0.041$.  From
the errors on $\tau(\nu)$ in the bottom panel, we estimate that a
component that has its width and center should not have $\tau_0 \gtrsim
0.02$.  This is smaller than the 0.041 that we just calculated.  Since
$T_k$ smaller than 1000 K would produce an even larger peak opacity, the
true $T_k$ cannot be far below 1000 K; moreover, we could probably
include this component in the opacity fit, which would slightly reduce
the $\tau_0$ for the other components (and there would be significant
uncertainties). 

\subsubsection{ 3C315}

\begin{figure}[p!] 
\begin{center}
\includegraphics[height=6in] {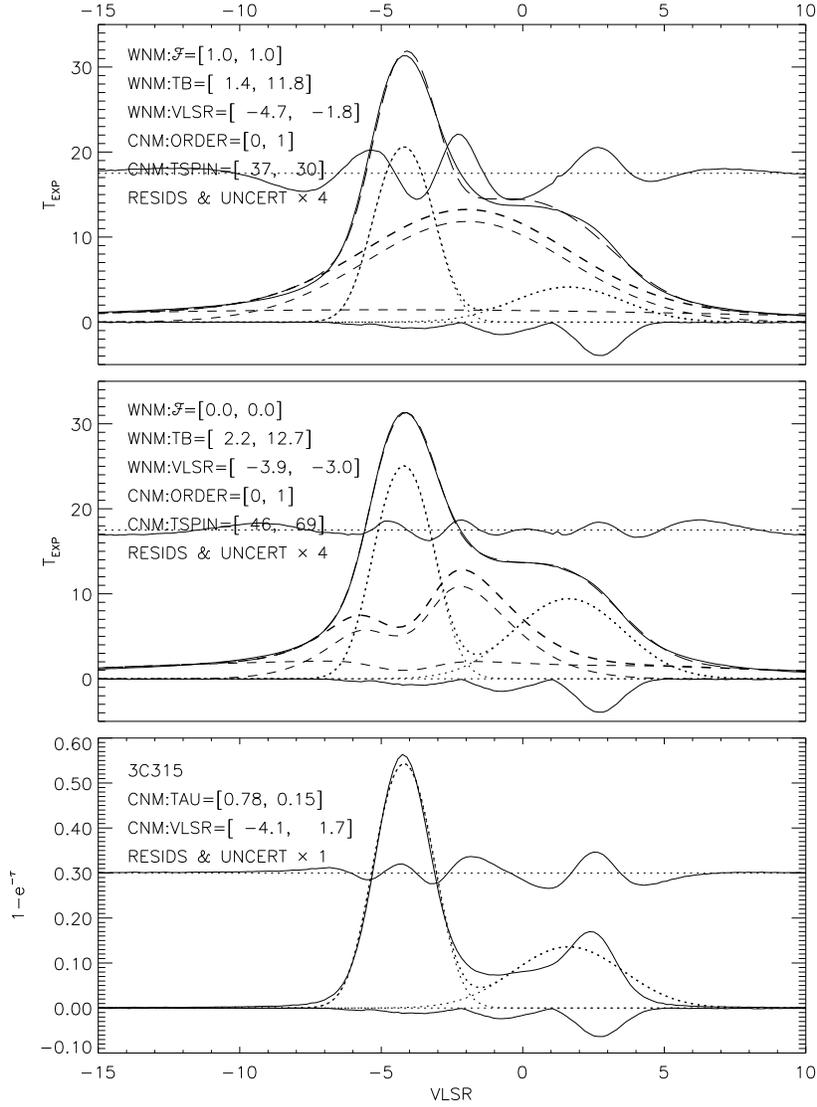} 

\caption{Data for 3C315.  Top two panels: data and fits having largest
(top) and smallest residuals for the trial fits to $T_{exp}(\nu)$. 
Bottom panel: data and fits for the opacity spectrum.  For a more
detailed description of this figure, see \S\ref{unusualsources}.  Here,
the top two panels are similar to the top panel in Figure
\ref{zec20plot}, and the bottom panel is like the bottom panel in that
figure. \label{3C315abplot} }

\end{center}
\end{figure}

	Figure \ref{3C315abplot} exhibits the results for 3C315.  The
opacity spectrum is fit to within the uncertainties by two CNM Gaussians
that overlap only slightly.  The emission spectrum is very well fit by
two WNM Gaussians with nearly equal central velocities; one is wide and
weak and provides a floor over which the the narrower, more intense WNM
Gaussian sits.  The residuals and uncertainties are expanded by a factor
of four on the top two panels.  The differing residuals are obvious.  As
for 3C120, the residuals are affected almost exclusively by ${\cal F}$,
here because of the small overlap of the two CNM components. 

	The intense, narrow WNM Gaussian has halfwidth $6.6 \pm 1.2$ km
s$^{-1}$, corresponding to $T_{kmax} < 950$ K.  From Table
\ref{bigtable}, its peak intrinsic weighted mean $T_{0,k} = 12.5 \pm
0.4$ K, so that its peak opacity $\tau_0 \geq (12.5/950)= 0.013$.  This
is somewhat smaller than the $\sim 0.02$ that one visually estimates
from the bottom panel for the upper limit for its opacity, so the
results are mutually consistent and its true temperature can be (but is
not necessarily) somewhat below 950 K, say $\sim 600$ K. 

\subsubsection{Commentary}

	Table \ref{worstbest} illustrates how the CNM spin temperatures
$T_s$ and the WNM intrinsic peak brightness temperatures $T_0$ change
from the worst and best fits, and also lists the fit adopted in our
standard Table \ref{bigtable}; these values are in the three pairs of
columns. We don't show the changes in $V_{LSR}$ or linewidth because
they are small. The adopted values from Table \ref{bigtable} are all
closer to the best case values than the worst case ones, and the
uncertainties on the adopted values look reasonable given the scatter of
the worst/best values. 

	Our main results in Table \ref{bigtable} are not those from the
particular ${\cal F}_k$ that yields the smallest variance. The reason is
that each derived variance is a result of an independent numerical
experiment and model fit. The results of each experiment should be
regarded as a set of independent estimates for the HI component
parameters. As with any set of independent measurements, the best
estimate for the derived parameters is an appropriately weighted average
of their independent estimates. We perform this weighting as described
in \S \ref{physcalc}.

	The WNM Gaussian centers and widths (and, to a lesser extent the
intensities) are not strongly affected by the choice of ${\cal F}$. This
is easy to understand. We can mentally divide the expected profile into
two portions, the ``center'' where the CNM absorption exists and and the
``outside'' where it does not. 

	Consider 3C120 above in Figure \ref{3C120abplot}. In both upper
panels, the WNM Gaussian parameters are determined mainly by the outside
portions: after all, it is to represent  these outside portions that the
WNM component is required. The opacity has no effect on these outside
portions; this is why the WNM Gaussian properties don't change much
between the two panels.

	In contrast, the central portion of the expected profile is
highly affected by ${\cal F}$. The central portion, both for 3C120 and
3C137, has narrow structure, and this structure can only come from the
narrower opacity components.  For these two sources, this narrower
structure is best represented by allowing the CNM components to absorb
the WNM emission from behind. This choice, with WNM behind, means that
the WNM contributes less to the emission and the CNM spin temperatures
must be larger to compensate.

\section{ COMPARISON OF GAUSSIAN-FIT RESULTS WITH SLOPE METHOD}

\label{slopevsgauss}

	Here we compare the two methods, both of which provide spin
temperatures using physically correct radiative transfer.  For each
method we provide two measurements of spin temperature for the two
extreme values of $\cal F$.  For ${\cal F}=0$ we designate the
slope-derived spin temperatures by the symbols $T_{s,S,{\cal F}=0}$ and
the Gaussian-derived ones with $T_{s,G,{\cal F}=0}$, with equivalent
designations for ${\cal F}=1$. 

	To obtain spin temperatures with the slope method (\S
\ref{mebold}),  one makes a ``slope plot'' by plotting $T_{exp}(\nu )$
versus $(1 - e^{-\tau(\nu)})$ and visually estimates the slope and the
intercept.  The intercept is $T_{E,WNM}$.  For ${\cal F} = 1$ the slope
is $T_s$; for ${\cal F} = 0$ the slope is $T_s - T_{E,WNM}$ (equation
\ref{slopeeqn}).  Our technique of Gaussian fitting also provides $T_s$
and $T_{E,WNM}$.

	Before beginning, we note that our least-square fits produce
reasonable facsimiles of the observed $T_{exp}(\nu)$ and
$e^{-\tau(\nu)}$ profiles. Thus, the slope plot of the observed
$T_{exp}(\nu)$ versus $e^{-\tau(\nu)}$ is very similar to that of the
predicted one, and both  provide similar estimates of $T_{s,S}$.

	Nevertheless, because of the nonuniqueness of Gaussians it is
conceivable that the model embodied in our least squares fit bears no
relationship to the real situation for a particular source.  Even so,
our fit is a numerical expression of {\it some} physical situation that
happens to match the observed data, and our Gaussian parameters are
necessarily correct for this (perhaps fictitious) situation. Therefore,
if the slope method disagrees with the Gaussian parameters for this
situation, the slope method is necessarily incorrect for this possibly
fictitious situation. In fact, however, we believe that our Gaussian
fits are usually reasonably good approximations to the real situations
and a disagreement between $T_{s,S}$ and $T_{s,G}$ indicates a problem
with the slope method.

\subsection{ Two representative examples}

	We compare the results derived from the slope method and our
Gaussian fits for two representative example sources, one with unblended
opacity components and one with blended components.  We show the results
in three-panel plots (Figures \ref{3C207pubplot} and
\ref{P0531+19_pubplot}) in which the top two panels are the usual
ones described in \S \ref{twopanelplots}. The bottom panel shows the
slope plot of $T_{exp}(\nu)$ versus $\tau(\nu)$; the solid line is the
data and the dashed line the least-squares fit.  The heavy dashed lines
show the visually-estimated slopes. In the upper left of each bottom
panel are the ranges of $T_s$ allowed by the slopes and intercepts. 

\subsubsection{ Unblended opacity components}

\begin{figure}[p!] 
\begin{center} 
\includegraphics[height=6in]{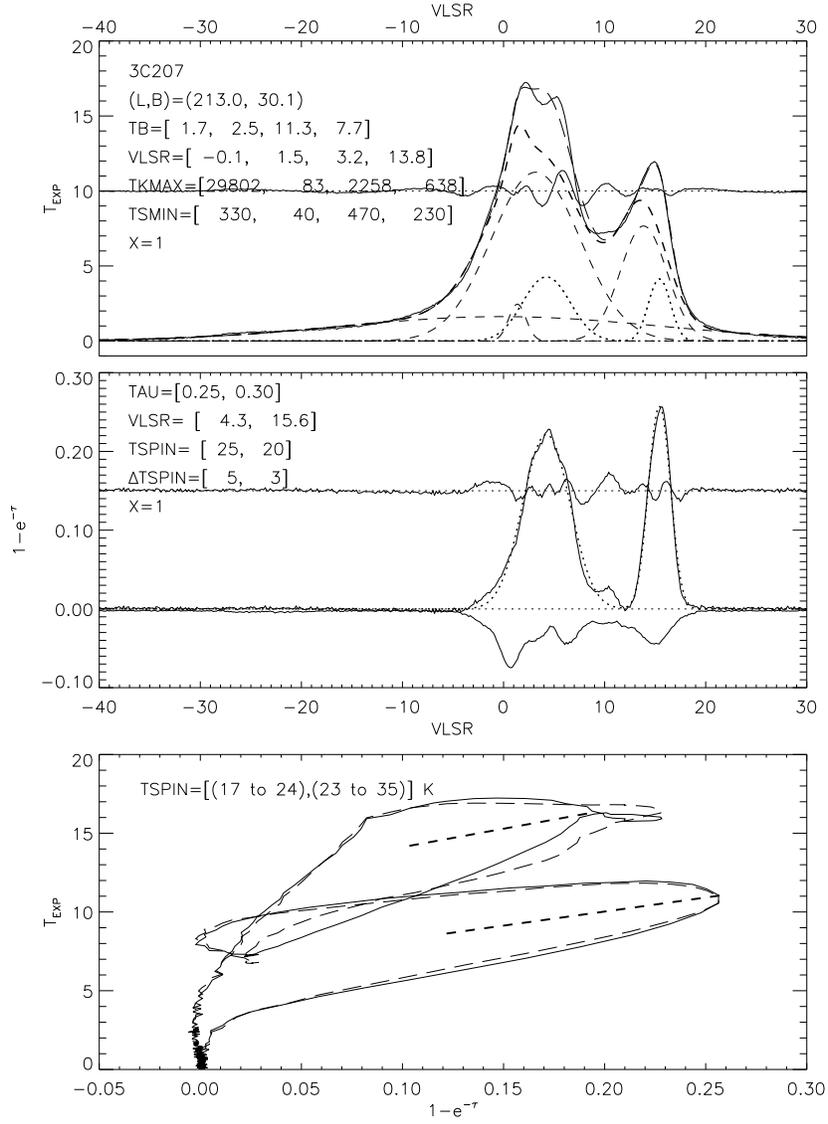} 

\caption{For 3C207, comparison of Gaussian-derived and slope-derived
spin temperatures for ${\cal F}=(0,1)$.  The top two panels are the same
as the two-panel plots such as Figure \ref{zec20plot}.  The bottom panel
is the slope plot $T_{exp}(\nu)$ versus $[1 - e^{-\tau(\nu)}]$ discussed
in \S \ref{slopevsgauss}.  \label {3C207pubplot} } \end{center}
\end{figure}

	Figure \ref{3C207pubplot} exhibits the data for 3C207.  There is
a pair of well-separated, relatively unblended CNM components, with two
associated strong WNM components that are slightly blended and a third
weak, narrow WNM component ($T_K < 100$ K) which is embedded in the
stronger WNM component.  Our CNM components have $T_s = (18 \pm 3, 18
\pm 5)$ K.  Each opacity component produces a well-defined lobe on the
slope plot; the slopes imply $\sim 17$ and 23 K if $F=1$, which is in
satisfactory agreement with our CNM components.  This result is typical:
in all cases with unblended CNM opacity components the slope method
works well, even with blended WNM components. 

\subsubsection{ Blended multicomponent opacity profiles}

\begin{figure}[p!]
\begin{center}
\includegraphics[height=6in] {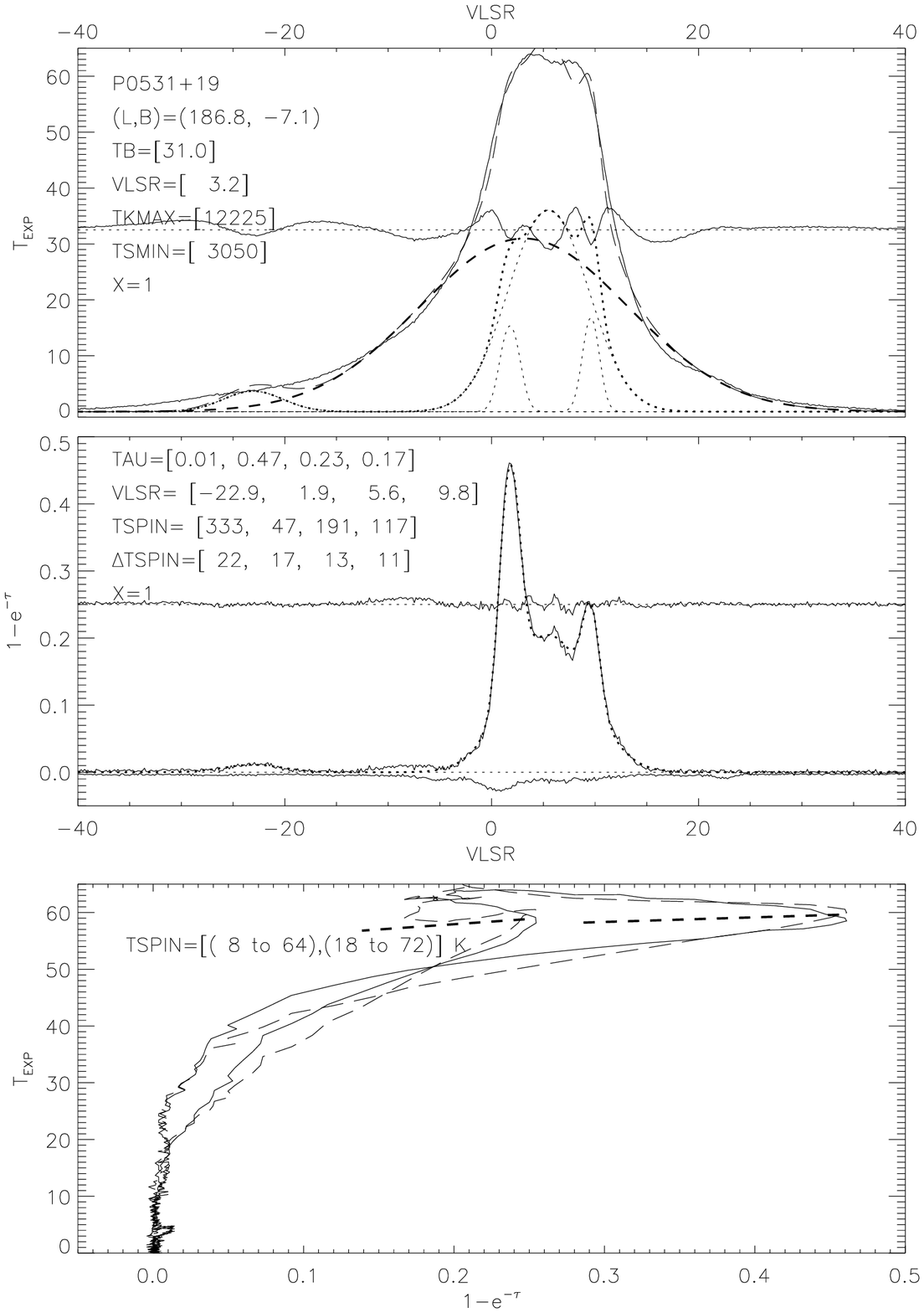}

\caption{For P0531+19, comparison of Gaussian-derived and slope-derived
spin temperatures for ${\cal F}=(0,1)$.  The top two panels are the same
as the two-panel plots such as Figure \ref{zec20plot}.  The bottom panel
is the slope plot $T_{exp}(\nu)$ versus $[1 - e^{-\tau(\nu)}]$ discussed
in \S \ref{slopevsgauss}.  \label {P0531+19_pubplot} }

\end{center}
\end{figure}

	Figure \ref{P0531+19_pubplot} exhibits the data for P0531+19. 
The opacity profile has three CNM components, two narrow ones with $T_s
= (40 \pm 2, 110 \pm 3)$ K for the strong and weak, respectively,
connected by a broad one with $T_s = (171 \pm 5)$ K.  The emission
profile consists of these plus a single broad WNM component.  The two
narrow CNM components produce well-defined lobes on the slope plot with
slopes of $\sim (8, 18)$ K for the strong and weak one, respectively. 
The broad CNM component overlaps both of the narrow components; it
contributes less opacity but as much or more emission than the narrow
components.  It does not produce a well-defined lobe, so that at the
velocity of each lobe on the slope plot both a narrow and the wide
component contribute. Thus each slope-derived spin temperature is in
some sense applicable to both its narrow and wide component. 

	The slope-derived temperatures are are much colder than the
Gaussian-derived ones.  Thus, the slope method doesn't work at all for
blended components.  It fails because the emission of a CNM component is
not proportional only to its opacity, but rather to the product of
opacity with spin temperature.  In this example, the spin temperature of
the broad opacity component is more than four times higher than that of
the stronger narrow component, so the broad component dominates the
emission while the narrow one dominates the opacity.  The emission of
the broad component doesn't change much over the narrow ones width, so
the slope is small. 

\subsection{ Sample of ten sources and conclusion}

	Table \ref{slopegauss} summarizes spin temperatures $T_{s,S}$
and $T_{s,G}$ for a set of ten representative sources.  The first four
sources have no overlapping CNM components.  The last six have
overlapping components, with one or more Gaussians contributing
significantly to the opacity at the velocity of each lobe in the slope
plot; the multiple values of $T_{s,G}$ for these components are listed
in square brackets.  The table lists temperatures for both ${\cal
F}=(0,1)$. 

	Sources like 3C207 with no overlapping components produce good
agreement between $T_{s,G}$ and $T_{s,S}$. But the agreement for sources
with overlapping components, like P0531+19, is abysmal.

	We conclude that the slope method works well when the opacity
profile is simple. However, with multiple components it does not work
well. We will proceed using our Gaussian fit results.

\section{SUMMARY} \label{summary}

	This paper discusses the observation and reduction techniques of
our large survey of the 21-cm line in emission and absorption. We use
Gaussian components and a simple but physically correct model to treat
the radiative transfer issues. The major topics are as follows.
\begin{enumerate}

	\item	\S \ref{stokespractice} presents the theory of
extracting the opacity and expected profiles from the on- and off-source
spectra. We apply this theory to the Arecibo data, which are
characterized by several effects common to most telescopes but amplified
at Arecibo because of its large sidelobes. The most serious instrumental
effect is the impossibility of getting a true off-source spectrum; we
develop an observing and reduction technique that not only solves this
problem but also provides reliable estimates of uncertainty for the
derived opacity and expected profiles. Our results compare well with
older data that are correct, but some older data are incorrect.
Surprisingly, stray radiation has little influence on Arecibo's emission
profiles (\S \ref{gaussianprocess}.

	\item	\S \ref{spintempderivation} discusses the radiative
transfer of the 21-cm line for the real case in which some of the gas is
Warm Neutral Medium (WNM) and some the Cold Neutral Medium (CNM). We
present a simple, physically correct model for this radiative transfer
for which we decompose the observed profiles into Gaussian components;
the CNM components are ordered along the line of sight so that some
absorb the emission of others, and the ensemble of CNM clouds is placed
an arbitrary fractional distance along the line of sight through the
WNM. 

	Because of our inclusion of radiative transfer, we derive spin
temperatures that are much lower than those from previous work. Our
temperatures are comparable to those derived for the Magellanic clouds
using the ``slope method''. The slope method is another simple,
physically correct model for the radiative transfer and works well for
simple profiles, but not multicomponent opacity profiles (\S
\ref{slopevsgauss}). 

\item Fitting Gaussians to spectra is a subjective and nonunique
process. \S \ref{gaussiancomps} devotes considerable discussion to our
method and process, with many illustrative examples to clarify our
subjective biases. For the opacity spectra, we generally fit the minimum
number of Gaussians required to reproduce them to within the
uncertainties, and for many sources the number of blended Gaussians is
small. For the expected emission profiles we fit a ``reasonable'' number
of additional WNM components (\S \ref{gaussianprocess}). We discuss the
effect of fitting either too few or too many Gaussians to a line profile
and conclude that the derived spin temperatures are not very much
affected.

	Some optical observers fit many Gaussians to reproduce line
shapes exactly. We argue that this procedure is not necessarily correct
because lines are always nonthermally broadened, in which case lines are
not necessarily Gaussians.

\item Paper II provides a detailed discussion of the WNM and CNM
properties, together with other astrophysical implications.

\end{enumerate}

\acknowledgements

	We acknowledge helpful discussions with Robert Braun, John
Dickey, Ed Jenkins, Dan Welty, and an unknown referee with an eagle eye.
This work was supported in part by NSF grants AST-9530590, AST-0097417,
and AST-9988341; and by the NAIC.


\eject

\begin{deluxetable}{c}
\tabletypesize{\scriptsize}
\tablewidth{0pc}
\tablecaption{Derived spectra for
all sourcees \label{fulldatatable} }
\tablehead{ \colhead{ This table is available only on-line as a 
machine-readable table}}
\startdata
FTP to vermi.berkeley.edu as anonymous and cd pub/zobs; get tab1.txt \\
This table should also be available on the electronic edition of Ap.J.
 \enddata
\end{deluxetable}

\eject

\begin{deluxetable}{ccccccccc}
\tabletypesize{\scriptsize}
\tablewidth{0pc}
\tablecaption{Table of Gaussian Fit Parameters \label{bigtable} }
\tablehead{
\colhead{ $T_B$ } &
\colhead{ $ \tau $ } &
\colhead{ $ V_{LSR} $ } &
\colhead{ $\Delta V$ } &
\colhead{ $ T_s$ } &
\colhead{ $ T_{kmax}$ } &
\colhead{ $ N(HI)_{20}$ } &
\colhead{ $ {\cal F}$ or O} &
\colhead{ $ (l/b/$SOURCE) }
}

\startdata
$    0.72\pm    0.04$ & $   0.004$ & $   -12.6\pm     1.0$ & $   53.58\pm    2.70$ &   180&  62746&  0.75 & $  0.50 \pm   0.29 $& 118.6/$-52.7$/3C18 \\
   16.71& $   0.669\pm   0.011$ & $    -8.9\pm     0.0$ & $    2.43\pm    0.03$ & $      34.25\pm   7.89$ &   129&  1.09&1&118.6/$-52.7$/3C18 \\
   19.80& $   0.077\pm   0.019$ & $    -6.7\pm     0.2$ & $    8.62\pm    0.70$ & $     267.80\pm   5.28$ &  1625&  3.46&0&118.6/$-52.7$/3C18 \\
    8.09& $   0.181\pm   0.018$ & $    -5.7\pm     0.1$ & $    3.98\pm    0.25$ & $      48.75\pm   9.41$ &   346&  0.69&2&118.6/$-52.7$/3C18 \\[4pt]
$    1.67\pm    0.12$ & $   0.002$ & $    -7.1\pm     0.4$ & $   26.21\pm    1.37$ &   830&  15018&  0.86 & $  0.50 \pm   0.29 $& 129.4/$-49.3$/3C33-1 \\
   10.41& $   0.034\pm   0.001$ & $    -4.6\pm     0.1$ & $    9.44\pm    0.30$ & $     309.63\pm   4.79$ &  1949&  1.95&0&129.4/$-49.3$/3C33-1 \\[4pt]
$    2.35\pm    0.19$ & $   0.011$ & $    -7.2\pm     0.5$ & $   22.25\pm    1.15$ &   210&  10822&  1.02 & $  0.50 \pm   0.29 $& 129.5/$-49.3$/3C33-2 \\
   10.20& $   0.059\pm   0.002$ & $    -4.2\pm     0.2$ & $    9.31\pm    0.42$ & $     177.85\pm   4.54$ &  1892&  1.90&0&129.5/$-49.3$/3C33-2 \\
\enddata

\tablecomments{For a complete description of table entries, see \S
\ref{tabular}. The ordering is by source name as follows: 3C, 4C, CTA, DW,
NRAO, P, T. CNM components have no error for $T_B$; WNM components
have no error for $\tau_0$. Temperatures are in K, velocities in km
s$^{-1}$, column densities in $10^{20}$ cm$^{-2}$. }

\end{deluxetable}



\begin{deluxetable}{cccccccc}
\tabletypesize{\footnotesize}
\tablewidth{0pc}
\tablecaption{Temperatures by two methods \label{slopegauss} }
\tablehead{
\colhead{Source}      & 
\colhead{$T_{s,S,{\cal F}=0}$} &
\colhead{$T_{s,G,{\cal F}=0}$} &
\colhead{$(T_{s,S}-T_{s,G})_{{\cal F}=0}$} &
\colhead{$|$} &
\colhead{$T_{s,S,{\cal F}=1}$} &
\colhead{$T_{s,G,{\cal F}=1}$} &
\colhead{$(T_{s,S}-T_{s,G})_{{\cal F}=1}$}
}

\startdata
3C333(${\cal F}=0.50$) & 29 & 25 & +4  &$|$ & \nodata & \nodata & \nodata \\
3C225b                 & 16 & 19 & -3  &$|$ & 11 & 16 & --5 \\
3C274.1                & 53 & 42 & +11 &$|$ & 48 & 36 & +12 \\
3C207                  & 24 & 24 & 0   &$|$ & 17 & 18 & --1 \\
3C207                  & 35 & 30 & +5  &$|$ & 23 & 19 & 4 \\
P0531+19    & 64 & 
$\left[ \begin{array}{cc} 72 \\ 206 \end{array} \right]$ & 
$\left[ \begin{array}{cc} -8 \\ -142 \end{array} \right]$  &$|$ & 8 & 
$\left[ \begin{array}{cc} 40 \\ 171 \end{array} \right]$  & 
$\left[ \begin{array}{cc} -32 \\ -163 \end{array} \right]$  \\[1.3pc]
P0531+19    & 72 & 
$\left[ \begin{array}{cc} 132 \\ 206 \end{array} \right]$  & 
$\left[ \begin{array}{cc} -60 \\ -134 \end{array} \right]$  &$|$ & 18 & 
$\left[ \begin{array}{cc} 110 \\ 171 \end{array} \right]$  & 
$\left[ \begin{array}{cc} -92 \\ -153 \end{array} \right]$  \\[1.3pc]
3C79        & 15 & 
$\left[ \begin{array}{cc} 15 \\ 147 \end{array} \right]$  & 
$\left[ \begin{array}{cc} 0 \\ -132 \end{array} \right]$   &$|$ & 0 & 
$\left[ \begin{array}{cc} 10 \\ 173 \end{array} \right]$  & 
$\left[ \begin{array}{cc} -10 \\ -173 \end{array} \right]$  \\
\enddata

\tablecomments{$T_{s,S}$ and $T_{s,G}$ are spin temperatures determined
from the slope and Gaussian methods, respectively.  We provide
comparisons for ${\cal F}=0$ in columns 2-4 and for ${\cal F}=1$ in
columns 5-7. Multiple values of $T_{s,G}$ enclosed in square brackets
mean that more than one Gaussian is associated with the single value of
$T_{s,S}$; see section \ref{slopevsgauss}.}

\end{deluxetable}


\end{document}